\documentclass[12pt,aps,showpacs]{revtex4}
\usepackage{amsmath,amssymb,amsfonts,slashed,epsfig}
\newcommand{\beeq}{\begin{equation}}
\newcommand{\eneq}{\end{equation}}
\newcommand{\be}{\begin{eqnarray}}
\newcommand{\ee}{\end{eqnarray}}
\newcommand{\bpic}{\begin{picture}}
\newcommand{\epic}{\end{picture}}
\newcommand{\bs}{\begin{scriptsize}}
\newcommand{\es}{\end{scriptsize}}

\def\dd{\partial}
\def\la{\raise.16ex\hbox{$\langle$} \, }
\def\ra{\, \raise.16ex\hbox{$\rangle$} }

\def\a{\alpha}

\def\l{\lambda}

\def\Box{\kern1pt\vbox{\hrule height 1.2pt\hbox{\vrule width 1.2pt\hskip 3pt
   \vbox{\vskip 6pt}\hskip 3pt\vrule width 0.6pt}\hrule height 0.6pt}\kern1pt}
\def\gtwid{\mathrel{\raise.3ex\hbox{$>$\kern-.75em\lower1ex\hbox{$\sim$}}}}
\def\ltwid{\mathrel{\raise.3ex\hbox{$<$\kern-.75em\lower1ex\hbox{$\sim$}}}}

\begin{document}


\title{Quantum corrected power spectra of\\ massless minimally coupled scalars during inflation:\\ Effects of Yukawa coupling versus quartic self-interaction}
\author{V. K. Onemli}\email{onemli@itu.edu.tr}

\affiliation{$^{\ast}$Department of Physics, Istanbul Technical
University, Maslak, Istanbul 34469, Turkey}
\begin{abstract}
We, in the first part, contemplate a massless minimally coupled scalar which is Yukawa-coupled to a massless Dirac fermion in a locally de Sitter background of an inflating spacetime. We compute the scalar's quantum corrected mode function, power spectrum, spectral index and the running of the spectral index at one-loop order. We find that the spectrum is slightly blue-tilted; hence, the amplitudes of fluctuations grow slightly toward the smaller scales. Then, in the second part, we apply the computation method used in the first part to a massless minimally coupled scalar with a quartic self-interaction in the same background and obtain exact analytic expressions for the associated quantities at one-loop order. In contrast to the Yukawa scalar, the spectrum in this case is slightly red-tilted; hence, the amplitudes of fluctuations grow slightly toward the larger scales.
\end{abstract}


\maketitle \vskip 0.2in \vspace{.4cm}

\section{Introduction}

Effectively massless and classically conformally noninvariant particles can exhibit enhanced quantum effects during inflation \cite{W1}. In the quantum vacuum, the virtual particle-antiparticle pairs that are produced during inflation endure longer and can become real. Endurance time is longest for the massless particles. In fact, long wavelength massless virtual particles can persist forever during inflation. Therefore, one would expect vast production of massless particles during inflation. Being massless, however, is not a sufficient condition for the particles to engender strong  quantum effects. For conformally invariant particles, the rate at which the virtual pairs emerge from the vacuum is suppressed by the inverse of the scale factor $a(t)$ and almost all massless particles are conformally invariant. (During inflation $a(t)\!=\!e^{Ht}$, where $H$ is the expansion rate and $t$ is the comoving time, hence the suppression is severe.) Massless minimally coupled (MMC) scalars and gravitons are the two exceptions known to be both massless and conformally noninvariant. They are, therefore, immensely produced and can engender strong quantum effects. The scalar~\cite{MC} and tensor perturbations~\cite{S} predicted by inflation indeed originate from the quantum fluctuations of these fields. They arise as tree order effects, because any virtual pair that pops up out of the vacuum is ripped apart by inflation and the constituents become real before they find time to annihilate each other, i.e., to close a loop in the Feynman diagram of the process. There is no one loop correction to the background after renormalization because the one loop diagrams are ultra-local. They consist of differentiated, coincident propagators. For example, in the MMC scalar model with a quartic self-interaction \cite{OW1,OW2} in a locally de Sitter background, our result for the one-loop vacuum expectation value of the stress-energy tensor is proportional to $\lim_{x'\rightarrow x} \partial_\mu \partial'_\nu i\Delta(x;x')$ which cannot involve an infrared logarithm. It is just a constant times the de Sitter metric. Any nonzero constant, however, must be absorbed into a renormalization of the bare cosmological constant, otherwise the initial expansion rate will not be $H$. Hence, it is only at two loops and higher that one can get the possibility of secular growth. At two loop order, MMC scalars with a quartic self-interaction yields~\cite{OW1,OW2,Completely} the violation of the weak energy condition causing the equation of state parameter $w\!<\!-1$ on cosmological scales and the gravitons slow inflation~\cite{TW1} as a back reaction.

In this paper, we consider a MMC scalar $\phi$ which is Yukawa-coupled to a massless Dirac fermion $\psi$. The Fermion affects the scalar through the self-mass squared.  Applying Schwinger-Keldish formalism~\cite{SK}, Duffy and Woodard showed~\cite{DW} that the scalar cannot quickly develop a large enough mass in this model. Hence, the particles are copiously produced during inflation. Massless fermions, on the other hand, are conformally invariant so their production rate is suppressed. Moreover, inflationary particle production of the Yukawa scalar derives the scalar field strength away from zero which affects the superhorizon fermions like a mass term and that further cuts off the fermion production.

Our interest, here, is to find the effect that the one-loop scalar self-mass squared has on the scalar's mode function and, in turn, on its power spectrum. To achieve these, we first perturbatively solve the linearized Schwinger-Keldish effective field equation that the scalar obeys at one-loop order applying the Green's function technique. Then, we employ the solution, i.e., the one-loop mode function, to obtain the scalar's quantum corrected power spectrum $\Delta^2_\phi$ from which we derive the spectral index $n_\phi$ and the running of the spectral index $\alpha_\phi$ at one-loop order.

Note that perturbation series in Yukawa theory goes like $c_\ell\!\left[\left(\!\frac{f^2}{4\pi^2}\!\right)\!\ln(a)\right]^\ell$ in the leading logarithm contribution at each perturbative order $\ell$. Here $f$ is the Yukawa coupling parameter and the constants $c_\ell$ are pure numbers which are assumed to be of order one. So perturbation theory breaks down when $\ln(a)\!\sim\!\!1/f^2$. For an arbitrarily weak coupling with $f\!\ll\!\!1$, the perturbation theory is valid for an arbitrarily long time. We assume that the Yukawa coupling is weak and apply the perturbation theory.

In Minkowski space pure Yukawa theory, it is well known that a loop of fermions does far more than just change the scalar self-mass squared; it renders the scalar effective potential unbounded from below~\cite{StHaRe}. This implies rolling down of the scalar freely. In the standard model, this tendency to decay is controlled by the positive effective potential of the gauge bosons and the need to avoid a possible instability for {\it large} Yukawa coupling of the top quark provides~\cite{FoJaJo} a constraint on the Higgs mass. The case is somewhat different in an inflating universe. Like in flat spacetime, a loop of fermions does change the scalar effective potential and from its large field expansion~\cite{MiWo} one can see that it is unbounded from below. Unlike in flat spacetime, however, the Hubble friction retards the scalar's rolling down its potential in an inflating spacetime. For {\it small} $f$ the effective potential's curvature is slight. Hence, the scalar's evolution is driven entirely by the pressure of the inflationary particle production. Only if the scalar's field strength can approach nonperturbatively large values does the unbounded effective potential---unless counteracted by some other effect as in the flat space standard model---begin to dominate.

Once the computation method used to study the Yukawa-coupled scalar is employed in the MMC scalar model with $\lambda\varphi^4$ self-interaction, exact analytic expressions for the scalar's quantum corrected mode function, power spectrum $\Delta^2_\varphi$, spectral index $n_\varphi$ and the running of the spectral index $\alpha_\varphi$ are obtained. Expansions of the analytical results we get in this paper agree with the series representations of the associated quantities we obtained in Ref.~\cite{VKO1}.

The outline of the paper is as follows. In Sec.~\ref{sec:Model} we describe the
background spacetime and the model. In Sec.~\ref{sec:EffectFieldEq}
we present the linearized one-loop Schwinger-Keldish  effective field
equation for a massless minimally coupled scalar which is Yukawa-coupled to a massless Dirac fermion in a locally de Sitter background first obtained in Ref.~\cite{DW}. We
solve it exactly in Sec.~\ref{sec:quantcorrectmodeYukawa}. We calculate
the scalar's one-loop corrected power spectrum, the spectral index
and the running of the spectral index
in Sec.~\ref{sec:Power}. In Sec.~\ref{sec:MMCS} we employ the computations done in Secs.~\ref{sec:quantcorrectmodeYukawa} and \ref{sec:Power} to obtain analytical expressions for the associated quantities for a MMC scalar with a quartic self-interaction during inflation. We discuss the implications of our results in Sec.~\ref{sec:conc}. The Appendixes include several key steps
of the computations.

\section{The Model}
\label{sec:Model}
Our background metric $g_{\mu\nu}$ describes an open
conformal coordinate patch of de~Sitter spacetime. The invariant
line element can be given in conformal and comoving
coordinates~as\beeq ds^2\!=\! g_{\mu\nu} dx^{\mu}
dx^{\nu}\!=\!a^2(\eta) \Bigl[ -d\eta^2 \!+\! d\vec{x} \cdot
d\vec{x} \Bigr]\!=\!-dt^2 \!+\! a^2(t) d\vec{x} \cdot d\vec{x}
\; , \label{mtrc}\eneq where the conformal factor \beeq a(\eta)\!=\!-{1 \over H \eta}\!=\!e^{H t}\!=\!a(t) \;  \label{cnfrml}\eneq
is normalized to $a\!=\!1$ when the state is
released in Bunch-Davies vacuum at conformal~time $\eta
\!=\!\eta_i\!\equiv\!-H^{-1}$ which corresponds to comoving time
$t\!=\!0$. The infinite future
$t\!\rightarrow\!\infty$, therefore, corresponds to $\eta\!\rightarrow\!0^-$. In our
notation, spacetime indices $\mu,\nu\!=\!0, 1, 2,\ldots, D\!-\!1$, spacetime coordinates $x^\mu\!=\!(x^0,\vec x)$ where  $x^0\!\equiv\!\eta$ and the partial derivatives $\partial_\mu\!=\!(\partial_0,\vec\nabla)$.

We consider a MMC real scalar which is Yukawa-coupled to a massless Dirac fermion in the background we just described. We follow the conventions given in Ref.~\cite{DW}. The Lagrangian density, in terms of unrenormalized scalar field
$\Phi(x)$ and Dirac field $\Psi_i(x)$, where the spinor index $i$ runs over $1,2,3,4$, is
\be
\mathcal{L}\!=\! -\frac12 \partial_{\mu} \Phi \partial_{\nu} \Phi
g^{\mu\nu} \sqrt{-g} \!+\! \overline{\Psi} e^{\mu}_{~b} \gamma^b \Big(i
\partial_{\mu} \!-\! \frac12 A_{\mu cd} J^{cd} \Big) \Psi \sqrt{-g} \!-\! \frac12 \xi_0 \Phi^2 R \sqrt{-g} \!-\! f_0 \Phi \overline{\Psi}
\Psi \sqrt{-g} \; .
\ee
Here, $g$ denotes the determinant of the metric and $R$ is the curvature scalar. The Latin letters $b, c, d$ denote the local Lorentz indices which run as $0, 1, 2,\ldots, D\!-\!1$.
The gamma matrices
$\gamma^{b}_{~ij}$ anti-commute
\beeq\{\gamma^b,\gamma^c\}\!\equiv\!\gamma^b\gamma^c\!+\!\gamma^c\gamma^b\!=\!-2 \eta^{bc} I\; .\eneq
The vierbein field $e_{\mu b}(x)$ relates the metric tensor of the spacetime to the Minkowski metric of the tangent space: $g_{\mu\nu}(x)\!=\!e_{\mu b}(x)
e_{\nu c}(x) \eta^{bc}$. The vierbein's vector index, denoted with a Greek letter, is raised and lowered by
the spacetime metric, e.g., $e^{\mu}_{~b}\!=\!g^{\mu\nu} e_{\nu b}$, whereas its local Lorentz
index is raised and lowered by the Minkowski metric, e.g., $e_{\mu}^{~b}\!=\!\eta^{bc} e_{\mu c}$.
Hence, $e^{\mu b}e_{\mu c}\!=\!\delta^{b}_{~c}$, $e^{\mu b}e_{\nu b}\!=\!\delta^{\mu}_{~\nu}$ and $g_{\mu\nu}\!=\!e_{\mu b}e_{\nu}^{~b}$. We use torsion free, metric compatible ($g_{\rho\sigma ;\mu}\!=\!0$, the semicolon denotes covariant derivative) connection
\beeq
\Gamma^\rho_{~\mu\nu}\!=\!\frac{g^{\rho\sigma}}{2}\left(g_{\sigma\mu,\nu}\!+\!g_{\nu\sigma,\mu}
\!-\!g_{\mu\nu,\sigma}\right)\; ,
\eneq
and vierbein compatible ($e_{\tau b ;\mu}\!=\!0$) spin connection
\beeq
A_{\mu cd}\!\equiv\! e^{\nu}_{~c} \left(e_{\nu d , \mu} \!-\! \Gamma^{\rho}_{~\mu \nu}
e_{\rho d}\right)\; .
\eneq
Choosing $e_{\mu b}\!=\!a
\eta_{\mu b}$ yields
\beeq
A_{\mu cd} \!=\! \left(\eta_{\mu c}
\partial_d\!-\!\eta_{\mu d} \partial_c \right) \ln(a) \; .
\eneq
The Lorentz representation
matrices
\beeq
J^{cd}\!\equiv\!\frac{i}{4} [\gamma^c,\gamma^d]\!=\!\frac{i}{4}(\gamma^c\gamma^d\!-\!\gamma^d\gamma^c)
\; .\eneq
The $\overline{\Psi} \!\equiv\! \Psi^{\dagger} \gamma^0$ is the Dirac
adjoint. The $\xi_0$ and $f_0$ denote the bare conformal and bare Yukawa couplings, respectively.

Introducing the renormalized fields
\beeq
\phi\!\equiv\!\frac{\Phi}{\sqrt{Z}} \;\; {\rm and} \;\; \psi\!\equiv\!\frac{\Psi}{\sqrt{Z_2}}\; ,
\eneq
and defining the field strength $(\delta Z, \delta Z_2)$, Yukawa $(\delta f)$ and conformal $(\delta \xi)$ counterterms,
\be
Z \equiv 1 \!+\! \delta Z , \; Z_2 \!\equiv\! 1 \!+\! \delta Z_2 , \;
\sqrt{Z} Z_2 f_0 \!\equiv\! f \!+\! \delta f , \; Z \xi_0 \!\equiv\! 0 \!+\! \delta \xi
\ee
bring the Lagrangian density to the form,
\be
&&\hspace{-1.5cm}{\cal L} \rightarrow -\frac12 a^{D-2} \partial_{\mu} \phi
\partial^{\mu} \phi \!+\! \Bigl( a^{\frac{D-1}2}
\overline{\psi} \Bigr) i \slashed\partial \Bigl(
a^{\frac{D-1}2} \psi \Bigr)\!\!-\!f a^D \phi \overline{\psi} \psi
\!-\!\frac12 \delta Z a^{D-2} \partial_{\mu} \phi \partial^{\mu} \phi\nonumber\\
&&\hspace{-1.3cm}+\frac12 \delta \xi (D\!-\!1) a^{D-2} \Bigl(2 \frac{\partial^2 a}{a} \!+\!
(D\!-\!4) \frac{\partial_{\mu} a}{a} \frac{\partial^{\mu} a}{a} \Bigr)
\phi^2 \!+\! \delta Z_2 \Bigl( a^{\frac{D-1}2} \overline{\psi} \Bigr) i
\slashed\partial \Bigl( a^{\frac{D-1}2} \psi
\Bigr)\!\!-\!\delta f a^D \phi \overline{\psi} \psi \; . \label{cterms}
\ee
The Dirac slashed derivative $\slashed\partial\!\equiv\!
\gamma^{\mu} \partial_{\mu}$. Note that, in this form of the Lagrangian density, the indices are raised and lowered by the Lorentz metric, e.g., $\partial^{\mu}\!=\!\eta^{\mu\nu} \partial_\nu$.

\section{Linearized Schwinger-Keldish Effective Field Equation}
\label{sec:EffectFieldEq}
In cases for which the ``in'' and ``out'' vacua differ, as in cosmology where the background is time dependent, the in-out matrix elements computed applying the usual Feynman formalism are not true expectation values.
The Schwinger-Keldish formalism~\cite{SK} is a covariant extension of the Feynman's formalism which yields true expectation values.

Quantum processes induce the scalar a self-mass which must be calculated using the Schwinger-Keldish formalism in cosmology. The scalar self-mass squared modifies the dynamical field equation that the scalar must obey. It is this linearized effective field equation that one needs to solve to obtain the quantum corrected scalar mode function.
The renormalized one-loop Yukawa-coupled scalar self-mass-squared was computed in Ref.~\cite{DW} employing the Schwinger-Keldish formalism. It can be given as
\be
&&\hspace{-0.4cm}M^2(x;x')\!=\!-\frac{f^2 a a'}{64 \pi^3}\partial^6\!\left\{\theta(\Delta \eta \!-\! \Delta x)
\Bigl(\ln\Bigl[\mu^2 (\Delta \eta^2 \!\!-\! \Delta x^2)\Bigr]\!\!-\! 1\Bigr)\!
\right\}\!-\!\frac{f^2 a a'}{8 \pi^2} \ln(a a') \partial^2 \delta^4(x \!-\! x')\nonumber\\
&&\hspace{9cm}-\delta \xi_{\mbox{\tiny fin}} 6 a \partial^2 a \delta^4(x\!-\!x') \!+\! \mathcal{O}(f^4)\;,\label{MsqrdYuk}
\ee
where $\theta$ is Heaviside step function and the coordinate intervals $\Delta\eta\!\equiv\!\eta\!-\!\eta'$ and $\Delta x\!\equiv\!\|\vec{x}\!-\!\vec{x}\,'\|$. The parameter $\mu$ is the dimensional regularization mass scale, and $\delta \xi_{\mbox{\tiny fin}}$ is a finite, order $f^2$ contribution in the conformal counterterm $\delta \xi$. The $M^2(x;x')$ enters
the linearized Schwinger-Keldish effective field equation as a
source term integrated against the scalar, \beeq
\partial_{\mu} \Bigl(\sqrt{-g}
g^{\mu\nu} \partial_{\nu}
\phi(x)\Bigr)\!=\!\!\int_{\eta_i}^\eta\!\! d\eta'\!\int\!d^3x'
M^2(x;x')\phi(x')\; .\label{effectivefieldeqn} \eneq
The quantum corrected scalar mode function $\phi(x;\vec{k})$ is
the solution of this integro-differential
equation of the form \beeq
\phi(x;\vec{k})\!=\!\mathrm{g}(\eta,k) e^{i\vec{k}\cdot\vec{x}} \; ,\label{yukscalmodefnc}
\eneq
where $\vec{k}$ is the comoving wave vector and $k\equiv\|\vec{k}\|$ is the comoving wave number. We solve the effective field equation
perturbatively, expanding the amplitude as a series in powers of
the loop counting parameter $\frac{f^2}{4\pi^2}$,
\beeq \mathrm{g}(\eta,k)\!\equiv\!\!\sum_{\ell=0}^{\infty}
\!\left(\!\frac{f^2}{4\pi^2}\!\right)^{\ell}\!\!\mathrm{g}_{\ell}(\eta,k)\;
.\label{solpert} \eneq The self mass squared can also be expressed as\beeq M^2(x;x')\!=\!\!
\sum_{\ell=0}^{\infty}\!\left(\!\frac{f^2}{4\pi^2}\!\right)^\ell\!\!\mathcal{M}^2_\ell(x;x')\; .\label{massloopexp}\eneq
Substituting Eqs.~(\ref{yukscalmodefnc})-(\ref{massloopexp}) into
Eq.~(\ref{effectivefieldeqn}) we find \beeq a^2 \Bigl[\partial_0^2 \!+\! 2
Ha\partial_0 \!+\! k^2\Bigr] \mathrm{g}_{\ell}(\eta,k) \!=\!
-\!\sum_{n=0}^{\ell} \!\int_{\eta_i}^\eta \!\!\! d\eta' \!\! \int
\!\! d^3x' \mathcal{M}^2_n(x;x') \mathrm{g}_{\ell-n}(\eta',k) e^{i
\vec{k} \cdot (\vec{x}\,' -\vec{x})} \; .\label{primpert} \eneq
Converting it from conformal
time $\eta$ to comoving time $t$ yields \beeq
\Bigl[ \frac{\partial^2}{\partial t^2}\!+\! 3 H
\frac{\partial}{\partial t} \!+\! \frac{k^2}{a^2}\Bigr] \mathrm{g}_\ell(\eta,k) \!=\!
-\frac{1}{a^4}\! \sum_{n=0}^{\ell} \!\int_{\eta_i}^\eta \!\!\! d\eta'
\!\! \int \!\! d^3x' \mathcal{M}^2_n(x;x') \mathrm{g}_{\ell-n}(\eta',k)
e^{i \vec{k} \cdot (\vec{x}\,' -\vec{x})} \; . \label{perteq} \eneq The tree-order ($\ell\!=\!0$) term in sum~(\ref{massloopexp}) \beeq\mathcal{M}^2_0(x;x')\!=\!0\;
,\label{1e0}\eneq since the scalar is classically massless. Hence, the tree-order mode function obeys
\begin{equation}
\ddot{\mathrm{g}}_0 \!+\! 3 H \dot{\mathrm{g}}_0 \!+\! \frac{k^2}{a^2}\mathrm{g}_0\!=\!0\; , \label{treeeqn}
\end{equation}
where an over-dot denotes derivative with respect to $t$. In de Sitter background the solution is the Bunch-Davies mode function which takes a simple form in $D\!=\!4$,
\begin{equation}
\mathrm{g}_0(\eta,k)\Bigl\vert_{D=4} \!=\! \frac{H}{\sqrt{2 k^3}}\!\left(\!1 \!-\! \frac{ik}{H a}\right)
\!\exp\!\left[\frac{ik}{Ha}\right] \!=\! \frac{H}{\sqrt{2 k^3}} \Bigl(\!1 \!+\! i k
\eta\Bigr) e^{-i k \eta} \; . \label{g0}
\end{equation}
The one-loop ($\ell\!=\!1$) mode function obeys,
\begin{equation}
\ddot{\mathrm{g}}_1 \!+\! 3 H \dot{\mathrm{g}}_1 \!+\! \frac{k^2}{a^2}\mathrm{g}_1\!=\!-\frac{1}{a^4}\!\!\int_{\eta_i}^\eta\!\!\! d\eta' \mathrm{g}_0(\eta',
k)\!\!\int\!\!d^3x' \mathcal{M}^2_1(x;x')e^{-i \vec{k}
\cdot(\vec{x}-\vec{x}\,')}\!\equiv\!\mathcal{S}(\eta, k)
\label{Integrog1}\; .
\end{equation}
The one-loop term $\mathcal{M}^2_1(x;x')$ can be read off from Eq.~(\ref{MsqrdYuk}).
A convenient choice~\cite{DW} for the parameters \beeq
\mu\!=\!\frac{H(t_i)}{2}\!=\!\frac{H}{2} \;\; {\rm and} \;\; \delta
\xi_{\mbox{\tiny fin}}\!=\!\frac{f^2}{32 \pi^2} \; ,
\eneq
yields
\be
&&\hspace{-1cm}\mathcal{M}^2_1(x;x')\!=\!-\frac{ a a'}{16\pi}\partial^6\!\left\{\theta(\Delta \eta \!-\! \Delta x)
\Bigl(\ln\Bigl[\frac{H^2}{4} \Big(\!\Delta \eta^2 \!-\! \Delta x^2\Big)\Bigr]\!\!-\! 1\Bigr)\!
\right\}\!-\!\frac{a a'}{2} \ln(a a') \partial^2 \delta^4(x \!-\! x')\nonumber\\
&&\hspace{9.9cm}-\frac{3a}{4}\partial^2 a \delta^4(x\!-\!x')\; .
\ee Thus, the source term $\mathcal{S}(\eta, k)$ in Eq.~(\ref{Integrog1}) can be written~\cite{DW} as
\be
&&\hspace{-0.3cm}\mathcal{S}(\eta, k)\!=\!H^2(\mathrm{g}_0^* \!-\! \mathrm{g}_0)\!\ln(a) \!+\! H\dot{\mathrm{g}}_0
\!-\!H^2 \mathrm{g}_0 2 i k \eta\!-\!\Big\{\!H\dot{\mathrm{g}}_0\!+\!H^2 \mathrm{g}_0(\!1 \!-\!
i k \eta)\!\Big\}\frac{e^{2 i k {\Delta \eta}_i}}{a\!-\!1}\!+\!H^2 \mathrm{g}_0 \frac{e^{2 i k
{\Delta \eta}_i}}{2 (a\!-\!1)^2}\nonumber\\
&&\hspace{1.4cm}-\Bigg\{\!\frac{H\dot{\mathrm{g}}_0}{a\!-\!1}\!+\!H^2 \mathrm{g}_0 2\ln(1 \!+\! H\eta)\!+\!
\frac{H^2\mathrm{g}_0}{a\!-\!1}\!-\!\frac{H^2\mathrm{g}_0}{2(a\!-\!1)^2}\!\Bigg\}\!+\!\frac{H^2\mathrm{g}_0^*}{2}\!\!\int_{-2 k \eta}^{-2 k \eta_i}
\!\!\!\!\!\!\!\!\!\! dx\,\,\Bigl[ \frac{e^{ix} \!-\! 1}{x} \Bigr]\nonumber\\
&&\hspace{9.9cm}-\frac{H^2\mathrm{g}_0}{2} \!\!\int_0^{2 k {\Delta \eta}_i} \!\!\!\!\!\!\!\!\!\! dx\,\,
\Bigl[\frac{e^{ix} \!-\! 1}{x} \Bigr]\; ,\label{source}
\ee
where ${\Delta \eta}_i\!\equiv\!\eta\!-\!\eta_i$. The first term in Eq.~(\ref{source}) has an oscillatory behavior
\be
&&\hspace{-1.3cm}H^2(\mathrm{g}_0^*\!-\! \mathrm{g}_0)\!\ln(a)\!=\!2H^2\mathrm{g}_0(0, k)i\Bigg\{\!\frac{k}{Ha}\cos\!{\Big(\!\frac{k}{Ha}\Big)}\!-\!\sin\!{\Big(\!\frac{k}{Ha}\Big)}\!\Bigg\}\!\ln(a)\; ,
\ee
whereas the second and third terms decay exponentially in time
\beeq
H\dot{\mathrm{g}}_0
\!-\!H^2 \mathrm{g}_0 2 i k \eta\!=\!H^2\mathrm{g}_0(0, k)\Bigg\{\!\frac{2ik}{Ha}\!+\!\frac{k^2}{H^2a^2}\!\Bigg\}e^{\frac{ik}{Ha}}\; .
\eneq
The fourth term,
\be
&&\hspace{-1cm}-\Big\{\!H\dot{\mathrm{g}}_0\!+\!H^2 \mathrm{g}_0(\!1 \!-\!
i k \eta)\!\Big\}\frac{e^{2 i k {\Delta \eta}_i}}{a\!-\!1}
\!=\!-H^2\mathrm{g}_0(0, k)e^{\frac{ik}{H}}\frac{e^{\frac{ik}{Ha}[a-1]}}{a\!-\!1}\nonumber\\
&&\hspace{4.5cm}=\!-H^2\mathrm{g}_0(0, k)e^{\frac{ik}{H}}\Bigg\{\!\frac{1}{a\!-\!1}\!+\!\!\sum_{n=1}^\infty\! \frac{1}{n!}\!\left[\frac{ik}{Ha}\right]^n\!\!(a\!-\!1)^{n-1}\Bigg\}\; ,\label{expsinglediv}
\ee
which also decays exponentially in time, consists of a singular part that diverges on the initial value surface and a finite part that vanishes there. (At $t\!=\!0$, $a\!=\!1$.) This initial value divergence arises because the state wave functional is specified at a finite time ($t\!=\!0$), as in many interacting quantum field theory models in cosmology, and the initial free state is not corrected. Perturbative initial state corrections can absorb~\cite{Completely} the initial value divergences. (Note also that, because the divergent part vanishes exponentially in time, it is irrelevant for the late time behavior---our real interest---of the mode function, and hence, of the power spectrum.) The fifth term behaves similarly to the fourth term
\be
&&\hspace{-0.2cm}H^2 \mathrm{g}_0 \frac{e^{2 i k
{\Delta \eta}_i}}{2 (a\!-\!1)^2}\!=\!H^2 \mathrm{g}_0\frac{e^{\frac{2ik}{Ha}[a-1]}}{2(a\!-\!1)^2}
\!=\!\frac{H^2}{2}\mathrm{g}_0(0, k)e^{\frac{ik}{H}}\Bigg\{\!\!\left[\!1\!-\!\frac{ik}{Ha}\right]\!\frac{1}{(a\!-\!1)^2}
\!+\!\left[\frac{ik}{Ha}\!+\!\frac{k^2}{H^2a^2}\right]
\!\frac{1}{(a\!-\!1)}\nonumber\\
&&\hspace{8.3cm}+\!\left[\!1\!-\!\frac{ik}{Ha}\right]\!\sum_{n=2}^\infty\! \frac{1}{n!}\!\left[\frac{ik}{Ha}\right]^n\!\!(a\!-\!1)^{n-2}\Bigg\}\; .\label{expsquarediv}
\ee
It decays exponentially in time and can be decomposed into a part that is singular on the initial value surface and a part that vanishes there. The sixth term \beeq
-\Bigg\{\!\frac{H\dot{\mathrm{g}}_0}{a\!-\!1}\!+\!\!H^2 \mathrm{g}_0 2\ln(1 \!+\! H\eta)\!+\!
\frac{H^2\mathrm{g}_0}{a\!-\!1}\!-\!\frac{H^2\mathrm{g}_0}{2(a\!-\!1)^2}\!\Bigg\}\; ,\label{sxth}
\eneq
diverges on the initial value surface and decays in time. The two definite integrals in the seventh and eighth terms in Eq.~(\ref{source}) are of the same form
\beeq
\int\!\! dx
\Bigl[\frac{e^{ix}\!-\!1}{x}\Bigl]=\!{\rm{Ei}}(ix) \!-\!\ln(x)\; .\label{inteiz}
\eneq
The exponential integral function ${\rm{Ei}}(ix)$ with $x\!\geq\!0$ can be expressed as
\beeq
{\rm{Ei}}(\pm ix)\!=\!{\rm ci}(x)\!\pm\!i\!\left[\pi\!+\!{\rm si}(x)\right]\; ,\label{Ei}
\eneq
where the cosine and sine integral functions, respectively, are
\be
{\rm ci}(x) \!\!&\equiv&\!\! -\!\int_x^{\infty} \!\!dt {\cos(t)
\over t} \!=\! \gamma \!+\! \ln(x) \!+\!\!\int_0^x \!\!dt {\cos(t)
\!-\!1 \over t} \label{ciint}\\
 {\rm si}(x) \!\!&\equiv&\!\! - \!\int_x^{\infty} \!\!dt
{\sin(t) \over t} \!=\! -\frac{\pi}2
\!+\!\!\int_0^x \!\!dt {\sin(t) \over t} \; . \label{siint}
\ee
The power series and asymptotic expansions of the ${\rm ci}(x)$ and ${\rm si}(x)$ functions are given in Eqs.~(\ref{cipowerser}), (\ref{sipowerser}), (\ref{ciasym}) and (\ref{siasym}).

Employing Eq.~(\ref{Ei}) in Eq.~(\ref{inteiz}), we evaluate the first definite integral in Eq.~(\ref{source}) as
\be
\hspace{-0.8cm}\int_{-2 k \eta}^{-2 k \eta_i}
\!\!\!\!\!\!\!\!\!\! dx\,\,\Bigl[ \frac{e^{ix} \!-\! 1}{x} \Bigr]\!\!\!&=&\!\!{\rm ci}\Big(\frac{2k}{H}\Big)\!-\!{\rm ci}\Big(\!\frac{2k}{Ha}\Big)\!-\!\ln(a)\!+\!i\!\left[{\rm si}\Big(\frac{2k}{H}\Big)\!-\!{\rm si}\Big(\!\frac{2k}{Ha}\Big)\right]\; .\label{firstint}
\ee
One can directly see that on the initial value surface $(a\!=\!1)$ the above integral vanishes. In the late time limit though, it asymptotes to a constant value
\beeq
\hspace{-0.8cm}\lim_{\eta\rightarrow 0^-}\int_{-2 k \eta}^{-2 k \eta_i}
\!\!\!\!\!\!\!\!\!\! dx\,\,\Bigl[ \frac{e^{ix} \!-\! 1}{x} \Bigr]
\!\!=\!{\rm ci}\Big(\frac{2k}{H}\Big)\!-\!\ln\!\Big(\frac{2k}{H}\Big)\!-\!\gamma\!+\!i\!\left[{\rm si}\Big(\frac{2k}{H}\Big)\!+\!\frac{\pi}{2}\right] \; .\label{asyfirstdefint}
\eneq
We used power series expansions~(\ref{cipowerser}) and (\ref{sipowerser}) of the functions ${\rm ci}(x)$ and ${\rm si}(x)$ in obtaining Eq.~(\ref{asyfirstdefint}). Similarly, the second integral in Eq.~(\ref{source}),
\be
\hspace{-0.8cm}\int_0^{2 k {\Delta \eta}_i}
\!\!\!\!\!\!\!\!\!\! dx\,\,\Bigl[ \frac{e^{ix} \!-\! 1}{x} \Bigr]
\!\!\!&=&\!\!{\rm ci}\Big(\frac{2k}{H}\!\left(1\!-\!a^{-1}\right)\!\!\Big)\!-\!\ln\!\Big(\frac{2k}{H}\!\left(1\!-\!a^{-1}\right)\!\!\Big)
\!-\!\gamma\!+\!i\!\left[{\rm si}\Big(\frac{2k}{H}\!\left(1\!-\!a^{-1}\right)\!\!\Big)\!+\!\frac{\pi}{2}\right]\; ,\label{finalint}
\ee
where we used the facts $\lim_{x\rightarrow 0}\left[\ln(x)\!-\!{\rm ci}(x)\right]\!=\!-\gamma$ and ${\rm si}(0)\!=\!-\pi/2$ which also imply that integral~(\ref{finalint}) vanishes on the initial value surface. In the late time limit, it asymptotes to the same constant that integral~(\ref{firstint}) asymptotes,
\beeq
\hspace{-0.8cm}\lim_{\eta\rightarrow 0^-}\int_{0}^{2 k {\Delta \eta}_i}
\!\!\!\!\!\!\!\!\!\! dx\,\,\Bigl[ \frac{e^{ix} \!-\! 1}{x} \Bigr]\!\!=\! {\rm ci}\Big(\frac{2k}{H}\Big)\!-\!\ln\!\Big(\frac{2k}{H}\Big)\!-\!\gamma\!+\!i\!\left[{\rm si}\Big(\frac{2k}{H}\Big)\!+\!\frac{\pi}{2}\right] \; .
\eneq
Thus, the contributions of the seventh and eighth terms in  Eq.~(\ref{source}) that involve integrals~(\ref{firstint}) and (\ref{finalint}) cancel each other in the late time limit where $g\!=\!g^*\!=\!H/\sqrt{2k^3}$. Hence, they together contribute to the source $\mathcal{S}(\eta, k)$ neither on the initial value surface nor in the late time limit.

Inserting Eqs.~(\ref{expsinglediv})-(\ref{sxth}), (\ref{firstint}) and (\ref{finalint}) into Eq.~(\ref{source}) we reexpress the source term as
\beeq
\mathcal{S}(\eta, k)\!=\!\mathcal{S}_{0}(\eta, k)\!+\!\mathcal{S}_{1}(\eta, k)\; ,
\eneq
where \be
&&\hspace{-1.5cm}\mathcal{S}_{0}(\eta, k)\!=\!-\Bigg\{\!\frac{H\dot{\mathrm{g}}_0}{a\!-\!1}\!+\!H^2\mathrm{g}_0\!\left[2\ln\!\Big(\!1\!-\!\frac{1}{a}\Big)
\!+\!\frac{1}{a\!-\!1}\!-\!\frac{1}{2(a\!-\!1)^2}\right]\nonumber\\
&&\hspace{1cm}+H^2\mathrm{g}_0(0, k)e^{\frac{ik}{H}}\!\Bigg[\!\frac{1}{a\!-\!1}\!-\!\frac{1}{2(a\!-\!1)}\!\left(\!\frac{ik}{Ha}\!+\!\frac{k^2}{H^2a^2}\!\right)
\!-\!\frac{1}{2(a\!-\!1)^2}\!\left(\!1\!-\!\frac{ik}{Ha}\!\right)\!\Bigg]\!\Bigg\}\; ,\label{S0}
\ee
is composed of terms that diverge on the initial value surface ($a\!=\!1$) as
\be
&&\hspace{-1.4cm}\lim_{\eta\rightarrow \eta_i}\mathcal{S}_{0}(\eta, k)\!=\!-H^2\mathrm{g}_0(0, k)e^{\frac{ik}{H}}\!\Bigg\{\!2\!\ln\!\Big(\!1\!-\!\frac{1}{a}\Big)
\!+\!\frac{2}{a\!-\!1}\!-\!\frac{1}{(a\!-\!1)^2}\nonumber\\
&&\hspace{4cm}-\frac{ik}{H}\!\left[2\ln\!\Big(\!1\!-\!\frac{1}{a}\Big)
\!+\!\frac{3}{2(a\!-\!1)}\!-\!\frac{1}{(a\!-\!1)^2}\right]\!\!-\!\frac{k^2}{H^2}\!\left[\frac{3}{2(a\!-\!1)}\right]\!\!\Bigg\}\; ,
\ee
but decay exponentially in the late time limit, \beeq\lim_{\eta\rightarrow 0^-}\mathcal{S}_0(\eta, k)\!=\!0\; ,\eneq whereas
\be
&&\hspace{-1cm}\mathcal{S}_{1}(\eta, k)\!=\!H^2(\mathrm{g}_0^* \!-\! \mathrm{g}_0)\!\ln(a) \!+\! H\dot{\mathrm{g}}_0
\!+\!H^2 \mathrm{g}_0 2\Big(\!\frac{ik}{Ha}\Big)\nonumber\\
&&\hspace{-0.3cm}-H^2\mathrm{g}_0(0, k)e^{\frac{ik}{H}}\Bigg\{\!\frac{ik}{Ha}\!+\!\!\sum_{n=2}^\infty\! \frac{1}{n!}\!\left[\frac{ik}{Ha}\right]^n\!\!(a\!-\!1)^{n-2}
\!\left[a\!-\!\frac{3}{2}\!+\!\frac{ik}{2Ha}\right]\!\!\Bigg\}\nonumber\\
&&\hspace{-0.3cm}+\frac{H^2}2 \mathrm{g}_0^*\Bigg\{\!{\rm ci}\Big(\frac{2k}{H}\Big)\!-\!{\rm ci}\Big(\!\frac{2k}{Ha}\Big)\!-\!\ln(a)\!+\!i\!\left[{\rm si}\Big(\frac{2k}{H}\Big)\!-\!{\rm si}\Big(\!\frac{2k}{Ha}\Big)\right]\!\!\Bigg\}\nonumber\\
&&\hspace{-0.3cm}-\frac{H^2}2 \mathrm{g}_0\Bigg\{\!{\rm ci}\Big(\frac{2k}{H}\!\left(1\!-\!a^{-1}\right)\!\!\Big)\!-\!\ln\!\Big(\frac{2k}{H}\!\left(1\!-\!a^{-1}\right)\!\!\Big)
\!-\!\gamma\!+\!i\!\left[{\rm si}\Big(\frac{2k}{H}\!\left(1\!-\!a^{-1}\right)\!\!\Big)\!+\!\frac{\pi}{2}\right]\!\!\Bigg\}\; ,\label{S1}
\ee
is composed of terms that are finite on the initial value surface,
\be
&&\hspace{-1.4cm}\lim_{\eta\rightarrow \eta_i}\mathcal{S}_{1}(\eta, k)\rightarrow H^2\mathrm{g}_0(0, k)e^{\frac{ik}{H}}\!\Bigg\{\!
\frac{ik}{H}\!+\!\frac{k^2}{H^2}\frac{3}{4}\!+\!\frac{ik^3}{H^3}\frac{1}{4}\Bigg\}\; ,
\ee
and, decay exponentially---like $\mathcal{S}_0(\eta, k)$---in the late time limit,
\beeq\lim_{\eta\rightarrow 0}\mathcal{S}_1(\eta, k)\!=\!0\; .\eneq

As was pointed out earlier, $\mathcal{S}_0(\eta, k)$ is composed entirely of terms that diverge on the initial value surface and rapidly redshift like the first and second inverse powers of the scale factor. Those are the same sorts of divergences we encountered~\cite{OW1,OW2} and resolved~\cite{Completely} in the vacuum expectation value of the stress-energy tensor in the MMC scalar model with a quartic self-interaction $\lambda\varphi^4$ during inflation. There is zero chance that the true vacuum is free vacuum in our model. That would not matter if we had an infrared stable theory and start at $t\!\rightarrow\!-\infty$. In that case one could obtain the full interacting state by evolving the free state over infinite time, just as in the flat spacetime. We start at $t\!=\!0$ because the theory is infrared sick, which means that we have to put in the state corrections. (They drop out at late times, so the leading infrared effects are independent of them.) Loop computations necessarily require perturbative corrections to the initial free state. Neglecting to correct the initial state gives rise to effective field equations that diverge on the initial value surface and/or sums of terms that redshift like the inverse powers of the scale factor. Initial state corrections to the initial state wave functional can absorb the initial value divergences and the redshifting terms and cure the theory. (Note also that, because $\mathcal{S}_0(\eta, k)\!\rightarrow\!0$ exponentially rapidly, as was pointed out, it has no relevance for the effect that the self-mass squared has on the late time behavior---the real interest in cosmology---of the scalar mode function.) Thus, it is the one-loop equation
\be
\ddot{\mathrm{g}}_1 \!+\! 3 H \dot{\mathrm{g}}_1 \!+\! \frac{k^2}{a^2}\mathrm{g}_1\!=\!\mathcal{S}_1(\eta, k)\; , \label{integrog1source}
\ee
that we need to solve. And that is done in the next section.
\section{quantum-corrected Yukawa scalar mode function}
\label{sec:quantcorrectmodeYukawa}

The solution $\mathrm{g}_1(\eta,k)$ of
Eq.~(\ref{integrog1source}) can be written as an integral over
comoving time as \beeq
\mathrm{g}_1(\eta,k)\!=\!\int_0^t\!\! dt'
G(t,t';k)\,\mathcal{S}_1(\eta', k)\; ,
\label{g1}\eneq where the Green's function\beeq
G(t,t';k)=\frac{\theta(t\!-\!t')}{W(t',k)}\left[\mathrm{g}_0(\eta, k)\mathrm{g}_0^*(\eta',
k)\!-\!\mathrm{g}_0^*(\eta, k)\mathrm{g}_0(\eta', k)\right] \; .\label{Green}\eneq
The $W(t', k)$, in Eq.~(\ref{Green}), is the Wronskian \beeq W(t', k)\!=\!\dot{\mathrm{g}}_0(\eta',
k)\mathrm{g}_0^*(\eta', k)\!-\!\mathrm{g}_0(\eta', k)\dot{\mathrm{g}}_0^*(\eta',
k)\!=\!\frac{-i}{a^3(\eta')}\!=\!iH^3\eta'^3=-ie^{-3Ht'} \;
.\label{Wronskian}\eneq Making the change of variable
$t'=\ln(a(\eta'))/H$ and using Eqs.~(\ref{g0}), (\ref{Green}) and
(\ref{Wronskian}) in Eq.~(\ref{g1}), we find
\beeq
\mathrm{g}_1(\eta,k)\!=\!\frac{i}{H}\!\int_1^a\!\!\! da'
a'^2\Bigg[\mathrm{g}_0(\eta, k)\mathrm{g}_0^*(\eta',
k)\!-\!\mathrm{g}_0^*(\eta, k)\mathrm{g}_0(\eta', k)\Bigg]\mathcal{S}_1(\eta', k)\; .
\eneq
To evaluate the integral on the right side it is useful to brake source term~(\ref{S1}) into a sum of six pieces
\beeq
\mathcal{S}_1(\eta, k)\!\equiv\!\sum_{s=1}^6\mathcal{S}_{1,s}(\eta, k)
\eneq where
\be
&&\hspace{4.8cm}\mathcal{S}_{1,1}(\eta, k)\!\equiv\!H^2(\mathrm{g}_0^* \!-\! \mathrm{g}_0)\!\ln(a) \label{S11}\\
&&\hspace{4.8cm}\mathcal{S}_{1,2}(\eta, k)\!\equiv\! H\dot{\mathrm{g}}_0\label{S12}\\
&&\hspace{4.8cm}\mathcal{S}_{1,3}(\eta, k)\!\equiv\! H^2 \mathrm{g}_0 2\Big(\!\frac{ik}{Ha}\Big)\label{S13}\\
&&\hspace{-1.2cm}\mathcal{S}_{1,4}(\eta, k)\!\equiv\!-H^2\mathrm{g}_0(0, k)e^{\frac{ik}{H}}\Bigg\{\!\frac{ik}{Ha}\!+\!\!\sum_{n=2}^\infty\! \frac{1}{n!}\!\left[\frac{ik}{Ha}\right]^n\!\!(a\!-\!1)^{n-2}
\!\left[a\!-\!\frac{3}{2}\!+\!\frac{ik}{2Ha}\right]\!\!\Bigg\}\label{S14}\\
&&\hspace{-1.2cm}\mathcal{S}_{1,5}(\eta, k)\!\equiv\!\frac{H^2}2 \mathrm{g}_0^*\Bigg\{\!{\rm ci}\Big(\frac{2k}{H}\Big)\!-\!{\rm ci}\Big(\!\frac{2k}{Ha}\Big)\!\!-\!\ln(a)\!+\!i\!\left[{\rm si}\Big(\frac{2k}{H}\Big)\!\!-\!{\rm si}\Big(\!\frac{2k}{Ha}\Big)\right]\!\!\Bigg\}\label{S15}\\
&&\hspace{-1.2cm}\mathcal{S}_{1,6}(\eta, k)\!\equiv\!-\frac{H^2}2 \mathrm{g}_0\Bigg\{\!{\rm ci}\Big(\frac{2k}{H}\!\left(1\!-\!a^{-1}\right)\!\!\Big)\!\!-\!\ln\!\Big(\frac{2k}{H}\!\left(1\!-\!a^{-1}\right)\!\!\Big)
\!\!-\!\gamma\!+\!i\!\left[{\rm si}\Big(\frac{2k}{H}\!\left(1\!-\!a^{-1}\right)\!\!\Big)\!+\!\frac{\pi}{2}\right]\!\!\Bigg\}\label{S16}\; .
\ee
Hence, the one loop mode function
\beeq
\mathrm{g}_1(\eta,k)\!\equiv\!\sum_{s=1}^6\mathrm{g}_{1,s}(\eta,k)\!\label{g1asSUM}\; ,
\eneq
where
\be
\mathrm{g}_{1,s}(\eta,k)\!=\!\frac{i}{H}\!\int_1^a\!\!\! da'
a'^2\Bigg[\mathrm{g}_0(\eta, k)\mathrm{g}_0^*(\eta',
k)\!-\!\mathrm{g}_0^*(\eta, k)\mathrm{g}_0(\eta', k)\Bigg]\mathcal{S}_{1,s}(\eta', k)\; .
\ee
Let us start computing the $\mathrm{g}_{1,1}(\eta,k)$ which can be written as
\be
&&\hspace{-1.2cm}\mathrm{g}_{1,1}(\eta,k)\!=\!-iH\Bigg\{\!\Bigg[\mathrm{g}_0^*(\eta, k)\!\!\int_1^a\!\!\! da'
a'^2\ln(a')\Big(|\mathrm{g}_0(\eta', k)|^2\!-\!\mathrm{g}_0^2(\eta', k)\!\Big)\Bigg]\!\!+\!\!\Bigg[{\rm C.C.}\Bigg]\!\Bigg\}\label{g11intsilk}\\
&&\hspace{-1.2cm}=\!-i\mathrm{g}_0^2(0, k)H\Bigg\{\!\!\left[\mathrm{g}_0^*(\eta, k)\!\!\left(\!\int_1^a\!\!\!
da'\!\left(\!\!a'^2\!+\!\frac{k^2}{H^2}\!\right)\!\ln(a')\!-\!\!\int_1^a\!\!\!da'
\!\left(\!\!a'\!-\!\frac{ik}{H}\!\right)^2\!\!\!e^{\frac{2ik}{Ha'}}\!\ln(a')\!\right)\!\right]\!\!+\!\!\Bigg[{\rm C.C.}\Bigg]\!\Bigg\}\;
,\label{g11integrals}\ee
where C.C. denotes complex conjugation. Employing Eqs. (\ref{intg11int1}) and (\ref{intg11int2}) in Eq.~(\ref{g11integrals}) we find
\be
&&\hspace{-0.4cm}\mathrm{g}_{1,1}(\eta,k)\!=\!-i\mathrm{g}_0^2(0, k)H\Bigg\{\!\Bigg\{\!\mathrm{g}_0^*(\eta, k)\Bigg[\!\Bigg[\!\frac{a^3}{3}\!\Big(\!\ln(a)\!-\!\frac{1}{3}\Big)\!+\!\frac{k^2}{H^2}a\Big(\!\ln(a)\!-\!1\Big)
\!-\!\frac{a}{3}e^{\frac{2ik}{Ha}}\Bigg(\!a\Big(\!\ln(a)\!-\!\frac{1}{3}\Big)\nonumber\\
&&\hspace{-0.4cm}\times\Big(a\!-\!\frac{2ik}{H}\Big)
\!\!+\!\frac{k^2}{H^2}\!\Big(\!\ln(a)\!-\!\frac{7}{3}\Big)\!\Bigg)
\!+\!\frac{2}{3}\frac{ik^3}{H^3}\Bigg(\!\frac{\ln^2(a)}{2}\!+\!\!\Big(\!\ln(a)\!-\!\frac{7}{3}\Big)\!\Big({\rm ci}\Big(\!\frac{2k}{Ha}\Big)\!-\!\ln\!\Big(\!\frac{2k}{H}\Big)\!-\!\gamma\nonumber\\
&&\hspace{-0.4cm}+i\Big[\frac{\pi}{2}\!+\!{\rm si}\Big(\!\frac{2k}{Ha}\Big)\!\Big]\Big)\!\!-\!\frac{25}{9}\!\Bigg)\!-\!\frac{4}{3a}\frac{k^4}{H^4}{}_{3}\mathcal{F}_{3}
\!\left(\!\!1, 1, 1; 2, 2, 2; \frac{2ik}{Ha}\!\right)\!\!\Bigg]\!\!-\!\Bigg[a\rightarrow\! 1\Bigg]\!\Bigg]\!\Bigg\}\!+\!\Bigg\{{\rm C.C.}\Bigg\}\!\Bigg\}\; ,\label{g1mid}
\ee
where the power series representation of the generalized hypergeometric function ${}_{3}\mathcal{F}_{3}$ is given in Eq.~(\ref{hyp33}). Note that $\mathrm{g}_{1,1}(\eta,k)$ can also be given as\be
&&\hspace{-0cm}\mathrm{g}_{1,1}(\eta,k)\!=\!-i\mathrm{g}_0^2(0, k)H\Bigg\{\!\Bigg[ \mathrm{g}_0^*(\eta,
k)\Bigg\{\!a^3\frac{\ln(a)}{3}\!-\!\frac{a^3}{9}\!+\!\frac{1}{9}\!+\!\frac{k^2}{H^2}\Big(a\ln(a)
\!-\!a\!+\!1\Big)\!-\!\frac{ik^3}{H^3}\frac{\ln^2(a)}{3}\nonumber\\
&&\hspace{2cm}-\!\sum_{n \doteq\,
0}^\infty\frac{1}{n!}\left(\frac{2ik}{H}\right)^n\!\!
\Bigg(\!\left(\frac{a^{3-n}}{3\!-\!n}\ln(a)-\frac{(a^{3-n}\!-\!1)}{(3\!-\!n)^2}\right)\!-\!\frac{2ik}{H}\!
\left(\frac{a^{2-n}}{2\!-\!n}\ln(a)\!-\!\frac{(a^{2-n}\!-\!1)}{(2\!-\!n)^2}\right)\nonumber\\&&\hspace{6cm}
-\frac{k^2}{H^2}\!\left(\frac{a^{1-n}}{1\!-\!n}\ln(a)
\!-\!\frac{(a^{1-n}\!-\!1)}{(1\!-\!n)^2}\right)\!\Bigg)\!\Bigg\}\Bigg]\!\!+\!\!\Bigg[{\rm C.C.}\Bigg]\!\Bigg\}\;
, \label{exactg11asasum}\ee where the symbol $\doteq$ indicates that the terms that diverge for each $n$, as
$n$ runs from $0$ to $\infty$, are to be excluded. In other words, when $n\!=\!1$ in the sum
the terms in the third parentheses, when $n\!=\!2$ the terms in the second parentheses and when $n\!=\!3$ the terms in
the first parentheses are to be excluded. If we express it in powers of $\frac{k}{Ha}$, we see that
the lowest order correction comes in cubic
order\be &&\hspace{-0.4cm}\mathrm{g}_{1,1}(\eta,k)\!=\!-\mathrm{g}_0(0,
k)\frac{i}{9}\Bigg\{\!\!\left(\!\frac{k}{Ha}\!\right)^3\!\Bigg[\!\frac{2a^3}{9}\!-\!\ln^2(a)\!-\!\frac{2\ln(a)}{3}
\!-\!\frac{2}{9}\Bigg]
\!\!-\!\left(\!\frac{k}{Ha}\!\right)^5\!\Bigg[\!\frac{2a^5}{125}\!-\!\frac{a^{3}}{9}
\!+\!\frac{a^2}{5}\!-\!\frac{\ln^2(a)}{10}
\nonumber\\
&&\hspace{0.3cm}-\frac{11\ln(a)}{75}\!-\!\frac{118}{1125}\Bigg]
\!\!+\!\left(\!\frac{k}{Ha}\!\right)^7\!\Bigg[\!\frac{6a^7}{8575}\!-\!\frac{a^5}{125}
\!+\!\frac{3a^4}{140}\!-\!\frac{a^3}{36}
\!+\!\frac{a^2}{50}\!-\!\frac{\ln^2(a)}{280}\!-\!\frac{107\ln(a)}{14700}
\!-\!\frac{4901}{771750}\Bigg]\nonumber\\
&&\hspace{12cm}+\mathcal{O}\!\left(\!\Big(\!\frac{k}{Ha}\!\Big)^{9}\!\right)\!\!\Bigg\}
\; .\label{g11power}\ee Note that, the leading logarithm term $\mathrm{g}_0(0,k)\frac{i}{9}\left(\!\frac{k}{Ha}\!\right)^3\ln^2(a)$ of $\mathrm{g}_{1,1}$ is the putative late time result of Ref.~\cite{DW}. Next, we compute
\be
&&\hspace{-1.3cm}\mathrm{g}_{1,2}(\eta,k)\!=\!-i\mathrm{g}^2_0(0, k)\frac{k^2}{H}\Bigg\{\!\mathrm{g}_0(\eta, k)\!\!\int_1^a\!\!\! da'
\Big[1\!+\!\frac{ik}{Ha'}\Big]\!-\!\mathrm{g}_0^*(\eta, k)\!\!\int_1^a\!\!\! da'
\Big[1\!-\!\frac{ik}{Ha'}\Big]e^{\frac{2ik}{Ha'}}\!\Bigg\}\;
.\label{g2ints}\ee
Using integrals (\ref{intexpint1}), (\ref{intexpint2})
and Eq.~(\ref{Ei}) in Eq.~(\ref{g2ints}) we find
\be
&&\hspace{-1cm}\mathrm{g}_{1,2}(\eta,k)\!=\!-i\mathrm{g}^2_0(0, k)\frac{k^2}{H}\Bigg\{\!\mathrm{g}_0(\eta, k)\!\Bigg[\!a\!-\!1\!+\!\frac{ik}{H}\ln(a)\!\Bigg]\nonumber\\
&&\hspace{1.3cm}-\mathrm{g}_0^*(\eta, k)\!\Bigg[ae^{\frac{2ik}{Ha}}\!-\!e^{\frac{2ik}{H}}\!-\!\frac{ik}{H}\!\left[{\rm ci}\Big(\!\frac{2k}{Ha}\Big)\!-\!{\rm ci}\Big(\!\frac{2k}{H}\Big)\!+\!i\!\left(\!{\rm si}\Big(\!\frac{2k}{Ha}\Big)\!-\!{\rm si}\Big(\!\frac{2k}{H}\Big)\!\right)\right]\!\Bigg]\!\Bigg\}\;
.\label{g2mid}\ee
Employing the power series expansions of exponential, ${\rm ci}(x)$ and ${\rm si}(x)$ functions given in Eqs.~(\ref{cipowerser}) and (\ref{sipowerser}) yields
\be
&&\hspace{-0.3cm}\mathrm{g}_{1,2}(\eta,k)\!=\!-\mathrm{g}_0(0, k)\frac{1}{3}\Bigg\{\!\!\left(\!\frac{k}{Ha}\!\right)^2\!\left[\frac{a^2}{2}-\frac{3}{2}+a^{-1}\right]
\!\!+\!i\!\left(\!\frac{k}{Ha}\!\right)^3
\!\left[\frac{a^3}{3}\!-\!\ln(a)\!-\!\frac{1}{3}\right]\!\!-\!\left(\!\frac{k}{Ha}\!\right)^4\!\Bigg[\frac{3a^4}{20}\!-\!\frac{a^2}{4}\nonumber\\
&&\hspace{-0.3cm}+\frac{a^{-1}}{10}\Bigg]
\!\!-\!i\!\left(\!\frac{k}{Ha}\!\right)^5
\!\left[\frac{4a^5}{75}\!-\!\frac{a^3}{6}\!+\!\frac{a^2}{6}\!-\!\frac{\ln(a)}{10}\!-\!\frac{4}{75}\right]\!
\!+\!\left(\!\frac{k}{Ha}\!\right)^6\!\Bigg[\frac{a^6}{63}\!-\!\frac{3a^4}{40}\!+\!\frac{a^3}{9}
\!-\!\frac{a^2}{16}\!+\!\frac{1}{144}\!+\!\frac{a^{-1}}{280}\Bigg]
\nonumber\\
&&\hspace{12cm}+\mathcal{O}\!\left(\!\Big(\!\frac{k}{Ha}\!\Big)^{7}\right)\!\!\Bigg\} \; .\label{g12power}
\ee
The third contribution
\be
&&\hspace{-1.2cm}\mathrm{g}_{1,3}(\eta,k)\!=\!-\mathrm{g}_0^2(0, k)2k\Bigg\{\!\mathrm{g}_0(\eta, k)\!\!\int_1^a\!\!\!
da' a'\!\left[1\!+\!\frac{k^2}{H^2a'^2}\right]\!\!-\!\mathrm{g}^*_0(\eta, k)\!\!\int_1^a\!\!\!da' a'\!\left[1\!-\!\frac{ik}{Ha'}\right]^2\!\!\!e^{\frac{2ik}{Ha'}}\!\Bigg\}\; , \label{g3ints} \ee
is computed similarly. Using integrals (\ref{g13int1}), (\ref{g13int2}) and Eq.~(\ref{Ei}) in Eq.~(\ref{g2ints}) we find
\be
&&\hspace{-0.7cm}\mathrm{g}_{1,3}(\eta,k)\!=\!-\mathrm{g}^2_0(0, k)2k\Bigg\{\!\mathrm{g}_0(\eta, k)\!\Bigg[\frac{a^2}{2}\!-\!\frac{1}{2}\!+\!\frac{k^2}{H^2}\ln(a)\!\Bigg]\!\!-\!\mathrm{g}_0^*(\eta, k)\!\Bigg[a\!\left[\frac{a}{2}\!-\!\frac{ik}{H}\right]\!e^{\frac{2ik}{Ha}}
\!-\!\left[\frac{1}{2}\!-\!\frac{ik}{H}\right]\!e^{\frac{2ik}{H}}\nonumber\\
&&\hspace{5cm}-\frac{k^2}{H^2}\!\left[{\rm ci}\Big(\!\frac{2k}{Ha}\Big)\!-\!{\rm ci}\Big(\!\frac{2k}{H}\Big)\!+\!i\!\left(\!{\rm si}\Big(\!\frac{2k}{Ha}\Big)\!-\!{\rm si}\Big(\!\frac{2k}{H}\Big)\!\right)\right]\!\Bigg]\!\Bigg\}\;
.\label{g3mid}\ee Employing the power series expansions of exponential, ${\rm ci}(x)$ and ${\rm si}(x)$ functions yields
\be
&&\hspace{-0.4cm}\mathrm{g}_{1,3}(\eta,k)\!=\!\mathrm{g}_0(0, k)\frac{1}{3}\Bigg\{\!i\!\left(\!\!\frac{k}{Ha}\!\right)\!\!\Big[2a\!-\!3\!+\!a^{-2}\Big]
\!+\!i\!\left(\!\!\frac{k}{Ha}\!\right)^3
\!\Bigg[\!\frac{4a^3}{15}\!+\!a\!-\!2\ln(a)\!-\!\frac{7}{6}\!-\!\frac{a^{-2}}{10}\Bigg]
\!\!-\!\left(\!\!\frac{k}{Ha}\!\right)^4\!\Bigg[\!\frac{a^4}{6}\nonumber\\
&&\hspace{-0.4cm}-\frac{2a}{3}\!+\!\frac{1}{2}\Bigg]
\!\!-\!i\!\left(\!\!\frac{k}{Ha}\!\right)^5
\!\Bigg[\!\frac{12a^5}{175}\!-\!\frac{2a^3}{15}\!+\!\frac{a}{4}\!-\!\frac{\ln(a)}{5}\!-\!\frac{109}{600}
\!-\!\frac{a^{-2}}{280}\Bigg]
\!\!+\!\left(\!\!\frac{k}{Ha}\!\right)^6\!\Bigg[\!\frac{a^6}{45}\!-\!\frac{a^4}{12}
\!+\!\frac{4a^3}{45}\!-\!\frac{a}{15}\!+\!\frac{7}{180}\Bigg]
\nonumber\\
&&\hspace{11.8cm}+\mathcal{O}\!\left(\!\Big(\!\frac{k}{Ha}\!\Big)^{7}\right)\!\!\Bigg\} \; .\label{g13power}
\ee The fourth contribution is
\be
&&\hspace{-1.4cm}\mathrm{g}_{1,4}(\eta,k)\!=\!-i\mathrm{g}_0(0, k)He^{\frac{ik}{H}}\!\!\int_1^a\!\!\! da'\Big[\mathrm{g}_0(\eta, k) \mathrm{g}^*_0(\eta', k)\!-\!\mathrm{g}^*_0(\eta, k) \mathrm{g}_0(\eta', k)\Big]\nonumber\\
&&\hspace{4.6cm}\times\!\Bigg[\frac{ik}{H}a'\!+\!\sum_{n=2}^\infty\! \frac{1}{n!}\!\left[\frac{ik}{H}\right]^n\!\left[1\!-\!\frac{1}{a'}\right]^{n-2}\!\left[a'\!-\!\frac{3}{2}\!+\!\frac{ik}{2Ha'}\right]
\!\Bigg]\; .\label{g14ints}
\ee
The integrals in Eq.~(\ref{g14ints}) are evaluated in Eqs.~(\ref{intg14int1})-(\ref{intg14int6}). Using these results we obtain
\be
&&\hspace{-1.1cm}\mathrm{g}_{1,4}(\eta,k)\!=\!-i\mathrm{g}^2_0(0, k)\frac{H}{2}e^{\frac{ik}{H}}\Bigg\{\!\mathrm{g}_0(\eta, k)\Bigg\{\!\Bigg[\!\frac{ik}{H}\Bigg\{\!a\Big(\!a\!+\!\frac{ik}{H}\Big)e^{-\frac{ik}{Ha}}
\!-\!\frac{k^2}{H^2}{\rm{Ei}}\Big(\!\!-\!\frac{ik}{Ha}\!\Big)\!\Bigg\}\nonumber\\
&&\hspace{-1.1cm}-\!\sum_{n=2}^\infty\sum_{m=0}^{n-2}\frac{(-1)^m\left(\frac{ik}{H}\right)^{n-m}}{n(n\!-\!1)(n\!-\!2\!-\!m)!\,m!}
\Bigg\{\!\frac{ik}{H}\Big[3\Gamma\Big(m\!-\!1, \frac{ik}{Ha}\!\Big)\!+\!2\Gamma\Big(m, \frac{ik}{Ha}\!\Big)\!-\!\Gamma\Big(m\!+\!1, \frac{ik}{Ha}\!\Big)\Big]\nonumber\\
&&\hspace{-1.1cm}+\frac{k^2}{H^2}\Big[2\Gamma\Big(m\!-\!2, \frac{ik}{Ha}\!\Big)\!+\!2\Gamma\Big(m\!-\!1, \frac{ik}{Ha}\!\Big)\Big]\!\Bigg\}\Bigg]\!\!-\!\Bigg[a\!\rightarrow\!1\Bigg]\!\Bigg\}\!-\!g^*_0(\eta, k)\Bigg\{\!\Bigg[\!\frac{ik}{H}\Bigg\{\!a\Big(\!a\!-\!\frac{ik}{H}\Big)e^{\frac{ik}{Ha}}\nonumber\\
&&\hspace{-1.1cm}-\frac{k^2}{H^2}{\rm{Ei}}\Big(\!\frac{ik}{Ha}\!\Big)\!\Bigg\}
\!+\!\sum_{n=2}^\infty\sum_{m=0}^{n-2}\frac{\left(\frac{ik}{H}\right)^{n-m}}{n(n\!-\!1)(n\!-\!2\!-\!m)!\,m!}
\Bigg\{\!\frac{ik}{H}\Big[3\Gamma\Big(m\!-\!1, -\frac{ik}{Ha}\!\Big)\!+\!4\Gamma\Big(m, -\frac{ik}{Ha}\!\Big)\nonumber\\
&&\hspace{0.1cm}+\Gamma\Big(m\!+\!1, -\frac{ik}{Ha}\!\Big)\Big]\!\!-\!\frac{k^2}{H^2}\Big[2\Gamma\Big(m\!-\!2, -\frac{ik}{Ha}\!\Big)\!+\!2\Gamma\Big(m\!-\!1, -\frac{ik}{Ha}\!\Big)\Big]\!\Bigg\}\Bigg]\!\!-\!\Bigg[a\!\rightarrow\!1\Bigg]\!\Bigg\}\!\Bigg\}\; ,\label{g4mid}
\ee where the incomplete gamma function $\Gamma(\alpha, z)$ is defined in Eq.~(\ref{incompgamma}) and its series expansion is given in Eq.~(\ref{seriesincompgamma}) when $\alpha$ is an integer. Power expanding the functions in Eq.~(\ref{g4mid}) yields
\be
&&\hspace{-0.4cm}\mathrm{g}_{1,4}(\eta,k)\!=\!-\mathrm{g}_0(0, k)\frac{1}{4}\Bigg\{\!i\!\left(\!\frac{k}{Ha}\!\right)\!\!\Bigg[\!\frac{4a}{3}\!-\!2\!+\!\frac{2a^{-2}}{3}\Bigg]
\!\!-\!\left(\!\frac{k}{Ha}\!\right)^2\!\!\Bigg[\!\frac{3a^2}{2}\!-\!3a\!+\!\frac{3}{2}\Bigg]
\!\!-\!i\!\left(\!\frac{k}{Ha}\!\right)^3
\!\!\Bigg[\!\frac{47a^3}{45}\!-\frac{7a^2}{3}\nonumber\\
&&\hspace{-0.5cm}+\frac{5a}{3}\!-\!\frac{4}{9}\!+\!\frac{a^{-2}}{15}\Bigg]
\!\!+\!\!\left(\!\frac{k}{Ha}\!\right)^4\!\!\Bigg[\!\frac{77a^4}{144}\!-\!\frac{5a^3}{4}\!+\!\frac{9a^2}{8}
\!-\!\frac{23a}{36}\!+\!\frac{11}{48}\Bigg]\!\!+\!i\!\left(\!\frac{k}{Ha}\!\right)^5
\!\!\Bigg[\!\frac{457a^5}{2100}\!-\!\frac{31a^4}{60}\!+\!\frac{23a^3}{45}\!-\!\frac{11a^2}{30}\nonumber\\
&&\hspace{-0.5cm}+\frac{11a}{60}\!-\!\frac{7}{225}\!+\!\frac{a^{-2}}{420}\Bigg]
\!\!-\!\left(\!\frac{k}{Ha}\!\right)^6\!\!\Bigg[\!\frac{53a^6}{720}\!-\!\frac{7a^5}{40}\!+\!\frac{49a^4}{288}
\!-\!\frac{139a^3}{1080}\!+\!\frac{13a^2}{144}\!-\!\frac{a}{24}\!+\!\frac{49}{4320}\Bigg]
\!\!+\!\mathcal{O}\!\left(\!\Big(\!\frac{k}{Ha}\!\Big)^{7}\right)\!\!\Bigg\}\; .\nonumber\\\label{g14power}
\ee The fifth contribution can be expressed as
\be
&&\hspace{-1.1cm}\mathrm{g}_{1,5}(\eta,k)\!=\!-i\mathrm{g}_0^2(0, k)\frac{H}{2}\Bigg\{\!\mathcal{A}(\eta, k)\!-\!\mathcal{B}(\eta, k)\!\Bigg\}
\; ,\label{intsing5}\ee
where we define\be
&&\hspace{-1.2cm}\mathcal{A}(\eta, k)\!\equiv\!\mathrm{g}_0^*(\eta, k)\!\!\!\int_1^a\!\!\!da'\!\left[{a'}^2\!\!+\!\frac{k^2}{H^2}\right]\!\!\Bigg\{\!{\rm ci}\Big(\frac{2k}{H}\Big)\!+\!i{\rm si}\Big(\frac{2k}{H}\Big)\!\!-\!\!\left[\ln(a')\!+\!{\rm ci}\Big(\!\frac{2k}{Ha'}\Big)
\!+\!i{\rm si}\Big(\!\frac{2k}{Ha'}\Big)\right]\!\!\Bigg\}\label{A}\; ,
\ee
and
\be
&&\hspace{-1.2cm}\mathcal{B}(\eta, k)\!\equiv\!\mathrm{g}_0(\eta, k)\!\!\!\int_1^a\!\!\!da'\!\left[a'\!+\!\frac{ik}{H}\right]^2\!\!\!e^{-\frac{2ik}{Ha'}}
\!\Bigg\{\!{\rm ci}\Big(\!\frac{2k}{H}\Big)\!+\!i{\rm si}\Big(\!\frac{2k}{H}\Big)\!\!-\!\!\left[\ln(a')\!+\!{\rm ci}\Big(\!\frac{2k}{Ha'}\Big)
\!+\!i{\rm si}\Big(\!\frac{2k}{Ha'}\Big)\right]\!\!\Bigg\}\; .\label{B}
\ee
Using Eqs.~(\ref{mathcalAint1})-(\ref{mathcalAint4}) in Eq.~(\ref{A}) and Eqs.~(\ref{mathcalBintfirst})-(\ref{mathcalBintthree}) in Eq.~(\ref{B}) we find
\be
&&\hspace{-0.7cm}\mathcal{A}(\eta, k)\!=\!\mathrm{g}_0^*(\eta, k)\Bigg\{\!\Bigg[\!\Big[{\rm ci}\Big(\frac{2k}{H}\Big)\!+\!i{\rm si}\Big(\frac{2k}{H}\Big)\Big]a\Big[\frac{a^2}{3}\!+\!\frac{k^2}{H^2}\Big]
\!\!-\!\frac{a^3}{3}\!\Big[{\rm ci}\Big(\!\frac{2k}{Ha}\Big)\!+\!\frac{1}{3}\cos\!\Big(\!\frac{2k}{Ha}\Big)\!+\!\ln(a)\!-\!\frac{1}{3}\Big]\nonumber\\
&&\hspace{-0.7cm}+\frac{k}{H}\frac{a^2}{9} \sin\!\Big(\!\frac{2k}{Ha}\Big)\!-\!\frac{k^2}{H^2}a\Big[{\rm ci}\Big(\!\frac{2k}{Ha}\Big)\!+\!\frac{7}{9}\cos\!\Big(\!\frac{2k}{Ha}\Big)\!+\!\ln(a)\!-\!1\Big]\!\!-\!\frac{k^3}{H^3}\frac{14}{9}\Big[{\rm si}\Big(\!\frac{2k}{Ha}\Big)\!+\!\frac{\pi}{2}\Big]\nonumber\\
&&\hspace{-0.2cm}-i\Bigg\{\!\frac{a^3}{3}\Big[{\rm si}\Big(\!\frac{2k}{Ha}\Big)\!+\!\frac{1}{3}\sin\!\Big(\!\frac{2k}{Ha}\Big)\Big]\!\!+\!\frac{k}{H}\frac{a^2}{9}
\cos\!\Big(\!\frac{2k}{Ha}\Big)\!+\!\frac{k^2}{H^2}a\Big[{\rm si}\Big(\!\frac{2k}{Ha}\Big)\!+\!\frac{7}{9}\sin\!\Big(\!\frac{2k}{Ha}\Big)\Big]\nonumber\\
&&\hspace{8.2cm}-\frac{k^3}{H^3}\frac{14}{9}{\rm ci}\Big(\!\frac{2k}{Ha}\Big)\!\Bigg\}\Bigg]\!\!-\!\Bigg[a\!\rightarrow\!1\Bigg]\!\Bigg\}\; ,\label{mathcalA}
\ee
and
\be
&&\hspace{-1.1cm}\mathcal{B}(\eta, k)\!=\!\mathrm{g}_0(\eta, k)\Bigg\{\!\Bigg[\!\Big[{\rm ci}\Big(\frac{2k}{H}\Big)\!-\!\ln\!\Big(\frac{2k}{H}\Big)\!-\!\gamma\!+\!i\Big\{{\rm si}\Big(\frac{2k}{H}\Big)\!+\!\frac{\pi}{2}\!\Big\}\Big]\frac{1}{3}\Big[e^{-\frac{2ik}{Ha}}
a\Big\{a^2\!+\!\frac{2ik}{H}a\!+\!\frac{k^2}{H^2}\Big\}\nonumber\\
&&\hspace{-1.2cm}+\frac{2ik^3}{H^3}{\rm{Ei}}\Big(\!\!-\!\frac{2ik}{Ha}\Big)\Big]\!\!+\!\frac{ik^3}{H^3}
\!\sum_{n=1}^\infty\frac{4}{n(2n)!}\Big[\Gamma\Big(\!2n\!-\!3,\frac{2ik}{Ha}\Big)
\!+\!\Gamma\Big(\!2n\!-\!2,\frac{2ik}{Ha}\Big)\!+\!\frac{1}{4}\Gamma\Big(\!2n\!-\!1,\frac{2ik}{Ha}\Big)\Big]\nonumber\\
&&\hspace{-1.2cm}+\frac{ik^3}{H^3}
\!\sum_{n=1}^\infty\!\frac{8}{(2n\!\!-\!\!1)(2n\!\!-\!\!1)!}\Big[\Gamma\Big(\!2n\!-\!4,\frac{2ik}{Ha}\Big)
\!\!+\!\Gamma\Big(\!2n\!-\!3,\frac{2ik}{Ha}\Big)
\!\!+\!\frac{1}{4}\Gamma\Big(\!2n\!-\!2,\frac{2ik}{Ha}\Big)\Big]\!\Bigg]\!\!-\!\!\Bigg[a\rightarrow\!1\Bigg]\!\Bigg\}\; .\label{mathcalB}
\ee
Expanding the functions in Eqs.~(\ref{mathcalA}) and~(\ref{mathcalB}) we obtain
\be
&&\hspace{-0.4cm}\mathrm{g}_{1,5}(\eta,k)\!=\!\mathrm{g}_0(0, k)\frac{1}{6}\Bigg\{\!i\!\left(\!\frac{k}{Ha}\!\right)\!\!\left[2a\ln(a)\!-\!\frac{8a}{3}\!+\!3\!-\!\frac{a^{-2}}{3}\right]
\!-\!\left(\!\frac{k}{Ha}\!\right)^2\!\Bigg[a^2\ln(a)\!-\!\frac{5a^2}{6}\!+\!\frac{3}{2}\!-\!\frac{2a^{-1}}{3}\Bigg]\nonumber\\
&&\hspace{-0.4cm}-i\!\left(\!\frac{k}{Ha}\!\right)^3
\!\left[\!\left(\!\frac{4a^3}{9}\!-\!a\!-\!\frac{14}{9}\!\right)\!\ln(a)\!-\!\frac{58a^3}{135}\!+\!\frac{10a}{3}\!-\!\frac{155}{54}
\!-\!\frac{a^{-2}}{30}\right]
\!\!+\!\left(\!\frac{k}{Ha}\!\right)^4\!\Bigg[\!\!\left(\!\frac{a^4}{6}\!-\!\frac{a^2}{2}\!-\!\frac{2a}{3}\!\right)\!\ln(a)\nonumber\\
&&\hspace{-0.4cm}-\frac{17a^4}{120}\!+\!\frac{17a^2}{12}\!-\!\frac{4a}{3}
\!+\!\frac{1}{8}\!-\!\frac{a^{-1}}{15}\Bigg]\!\!+\!i\!\left(\!\frac{k}{Ha}\!\right)^5
\!\Bigg[\!\!\left(\!\frac{4a^5}{75}\!-\!\frac{2a^3}{9}\!-\!\frac{a^2}{3}\!-\!\frac{a}{4}\!-\!\frac{7}{45}\!\right)\!\ln(a)
\!-\!\frac{143a^5}{2625}\!+\!\frac{89a^3}{135}\nonumber\\
&&\hspace{-0.4cm}-\frac{17a^2}{30}\!+\!\frac{a}{3}\!-\!\frac{9997}{27000}\!-\!\frac{a^{-2}}{840}\Bigg]\!
\!-\!\left(\!\frac{k}{Ha}\!\right)^6\!\Bigg[\!\!\left(\!\frac{2a^6}{135}\!-\!\frac{a^4}{12}\!-\!\frac{4a^3}{27}\!-\!\frac{a^2}{8}
\!-\!\frac{a}{15}\!\right)\!\ln(a)
\!-\!\frac{419a^6}{28350}\!+\!\frac{19a^4}{80}\!-\!\frac{68a^3}{405}\nonumber\\
&&\hspace{7.5cm}+\frac{5a^2}{48}\!-\!\frac{4a}{25}\!+\!\frac{11}{3240}\!-\!\frac{a^{-1}}{420}\Bigg]
\!\!+\!\mathcal{O}\!\left(\!\Big(\!\frac{k}{Ha}\!\Big)^{7}\right)\!\!\Bigg\}\; .\label{g15power}
\ee Similarly, the sixth contribution can be expressed as
\be
&&\hspace{-1.1cm}\mathrm{g}_{1,6}(\eta,k)\!=\!-i\mathrm{g}_0^2(0, k)\frac{H}{2}\Bigg\{\!\mathcal{C}(\eta, k)\!-\!\mathcal{D}(\eta, k)\!\Bigg\}
\; ,\label{intsing6}\ee
where\be
&&\hspace{-1.1cm}\mathcal{C}(\eta, k)\!\equiv\!\mathrm{g}_0(\eta, k)\!\!\int_1^a\!\!da'\!\left[{a'}^2\!\!+\!\frac{k^2}{H^2}\right]\!\!\Bigg\{\!{\rm ci}\Big(\frac{2k}{H}\Big[1\!-\!a'^{-1}\Big]\!\Big)
\!-\!\ln\!\Big(\frac{2k}{H}\Big[1\!-\!a'^{-1}\Big]\!\Big)\!-\!\gamma\nonumber\\
&&\hspace{9cm}+i\!\left[{\rm si}\Big(\frac{2k}{H}\Big[1\!-\!a'^{-1}\Big]\!\Big)\!+\!\frac{\pi}{2}\!\right]\!\!\Bigg\}\label{C}\; ,
\ee
and
\be
&&\hspace{-1.2cm}\mathcal{D}(\eta, k)\!\equiv\!\mathrm{g}^*_0(\eta, k)\!\!\int_1^a\!\!da'\!\left[a'\!-\!\frac{ik}{H}\right]^2\!\!\!e^{\frac{2ik}{Ha'}}
\!\Bigg\{\!{\rm ci}\Big(\frac{2k}{H}\Big[1\!-\!a'^{-1}\Big]\!\Big)
\!-\!\ln\!\Big(\frac{2k}{H}\Big[1\!-\!a'^{-1}\Big]\!\Big)\!-\!\gamma\nonumber\\
&&\hspace{9cm}+i\!\left[{\rm si}\Big(\frac{2k}{H}\Big[1\!-\!a'^{-1}\Big]\!\Big)\!+\!\frac{\pi}{2}\!\right]\!\!\Bigg\}\; .\label{D}
\ee Using Eqs.~(\ref{Cintone}) and (\ref{Cinttwo}) in Eq.~(\ref{C}) we get
\be
&&\hspace{-0.5cm}\mathcal{C}(\eta, k)\!\equiv\!\mathrm{g}_0(\eta, k)\Bigg\{\!\Bigg[\frac{a}{3}\Big[a\!+\!2\Big]\!\sin^2\!\Big(\frac{k}{H}\Big[1\!-\!a^{-1}\Big]\!\Big)
\!-\!\frac{2k}{3H}\Bigg\{\!\frac{a}{2}\sin\!\Big(\!\frac{2k}{H}\Big[\!1\!-\!a^{-1}\Big]\!\Big)\!-\!{\rm ci}\Big(\!\frac{2k}{Ha}\!\Big)\nonumber\\
&&\hspace{-0.6cm}\times\!\Big[\!\sin\!\Big(\!\frac{2k}{H}\Big)\!-\!\frac{k}{H}\cos\!\Big(\!\frac{2k}{H}\Big)
\!\Big]\!\!+\!\!\left[\frac{\pi}{2}\!+\!{\rm si}\Big(\!\frac{2k}{Ha}\!\Big)\right]\!\!\Big[\!\cos\!\Big(\!\frac{2k}{H}\Big)
\!+\!\frac{k}{H}\sin\!\Big(\!\frac{2k}{H}\Big)\!\Big]\!\!\Bigg\}
\!\!+\!\!\left[\frac{{a}^2}{3}\!+\!\frac{k^2}{H^2}\right]\!a\!\nonumber\\
&&\hspace{-0.6cm}\times\!\!\left[{\rm ci}\Big(\frac{2k}{H}\Big[\!1\!-\!a^{-1}\Big]\!\Big)
\!\!-\!\ln\!\Big(\frac{2k}{H}\Big[\!1\!-\!a^{-1}\Big]\!\Big)\!\!-\!\gamma\right]
\!\!+\!\!\left[\!\frac{1}{3}\!+\!\frac{k^2}{H^2}\right]\!\!\Bigg\{\!{\rm ci}\Big(\!\frac{2k}{Ha}\!\Big)\!\cos\!\Big(\!\frac{2k}{H}\Big)\!+\!\!\left[\pi\!+\!{\rm si}\Big(\!\frac{2k}{Ha}\!\Big)\right]\!\sin\!\Big(\!\frac{2k}{H}\Big)\nonumber\\
&&\hspace{-0.6cm}-{\rm ci}\Big(\frac{2k}{H}\Big[\!1\!-\!a^{-1}\Big]\!\Big)\!+\!\ln(1\!-\!a)\!\Bigg\}
\!\!-\!\frac{i}{3}\Bigg\{\!a\!\left[\frac{a}{2}\!+\!\!1\right]\! \sin\!\Big(\frac{2k}{H}\Big[\!1\!-\!a^{-1}\Big]\!\Big)\!+\!\!\left[\pi\!+\!{\rm si}\Big(\!\frac{2k}{Ha}\!\Big)\right]\!\cos\!\Big(\!\frac{2k}{H}\Big)\nonumber\\
&&\hspace{-0.6cm}-{\rm ci}\Big(\!\frac{2k}{Ha}\!\Big)\sin\!\Big(\!\frac{2k}{H}\Big)\!\!-\!\!\left[\!\frac{\pi}{2}\!+\!{\rm si}\Big(\frac{2k}{H}\Big[\!1\!-\!a^{-1}\Big]\!\Big)\right]
\!\!\left[a^3\!-\!1\!+\!\frac{k^2}{H^2}3(a\!-\!1)\right]\!\!+\!\!\frac{k}{H}\Bigg\{\!{\rm ci}\Big(\!\frac{2k}{Ha}\!\Big)\!\Big[2\cos\!\Big(\!\frac{2k}{H}\Big)\!\!-\!\frac{k}{H}\nonumber\\
&&\hspace{-0.6cm}\times\!\sin\!\Big(\!\frac{2k}{H}\Big)\Big]\!\!+\!\!\left[\!\frac{\pi}{2}\!+\!{\rm si}\Big(\!\frac{2k}{Ha}\!\Big)\right]\!\!\left[2\sin\!\Big(\!\frac{2k}{H}\Big)
\!\!+\!\!\frac{k}{H}\cos\!\Big(\!\frac{2k}{H}\Big)\right]\!\!-\!a\cos\!\Big(\frac{2k}{H}
\Big[\!1\!-\!a^{-1}\Big]\!\Big)\!\Bigg\}\!\Bigg\}\!\Bigg]\!\!-\!\!\Bigg[a\!\rightarrow\! 1\Bigg]\!\Bigg\}\; .\label{mathcalC}
\ee
We employ Eqs.~(\ref{anahalf1}) and (\ref{anahalf2}) in Eq.~(\ref{D}) and obtain
\be
&&\hspace{-0.3cm}\mathcal{D}(\eta, k)\!=\!\mathrm{g}^*_0(\eta, k)\!\Bigg\{\!\Bigg[\!\sum_{n=1}^\infty\!\sum_{m=0}^{2n}\!\frac{(-1)^n i^{1-m}\left(\frac{2k}{H}\right)^{2n-m+3}}{2n(2n\!-\!m)!\,m!}\Bigg\{\!\Gamma\Big(\!m\!-\!3,-\frac{2ik}{Ha}\Big)
\!+\!\Gamma\Big(\!m\!-\!2,-\frac{2ik}{Ha}\Big)\nonumber\\
&&\hspace{-0.3cm}+\frac{1}{4}\Gamma\Big(\!m\!-\!1,-\frac{2ik}{Ha}\Big)\!\Bigg\}
\!-\!i\!\sum_{n=1}^\infty\!\sum_{m=0}^{2n-1}\!\frac{(-1)^{n} i^{1-m}\left(\frac{2k}{H}\right)^{2n-m+2}}{(2n\!-\!1)(2n\!-\!1\!-\!m)!\,m!}
\Bigg\{\!\Gamma\Big(\!m\!-\!3,-\frac{2ik}{Ha}\Big)
\!+\!\Gamma\Big(\!m\!-\!2,-\frac{2ik}{Ha}\Big)\nonumber\\
&&\hspace{8.9cm}+\frac{1}{4}\Gamma\Big(\!m\!-\!1,-\frac{2ik}{Ha}\Big)\!\Bigg\}\Bigg]
\!\!-\!\!\Bigg[a\!\rightarrow\!1\Bigg]\!\Bigg\}\; .\label{mathcalD}
\ee
In powers of $\frac{k}{Ha}$, we have
\be
&&\hspace{-1.1cm}\mathrm{g}_{1,6}(\eta,k)\!=\!-\mathrm{g}_0(0, k)\frac{1}{6}\Bigg\{\!i\!\left(\!\!\frac{k}{Ha}\!\right)\!\!\Bigg[2a\ln(a)\!-\!\frac{8a}{3}\!+\!3\!-\!\frac{a^{-2}}{3}\Bigg]
\!\!-\!\left(\!\!\frac{k}{Ha}\!\right)^2
\!\Bigg[a^2\ln(a)\!-\!\frac{11a^2}{6}\!+\!3a\!-\!\frac{3}{2}\nonumber\\
&&\hspace{-1.2cm}+\frac{a^{-1}}{3}\Bigg]\!\!-\!i\!\left(\!\!\frac{k}{Ha}\!\right)^3
\!\Bigg[\!\!\left(\!\frac{4a^3}{9}\!-\!a\!-\!\frac{14}{9}\!\right)\!\ln(a)\!-\!\frac{148a^3}{135}\!+\!2a^2\!+\!\frac{4a}{3}
\!-\!\frac{119}{54}\!-\!\frac{a^{-2}}{30}\Bigg]\!\!+\!\!\left(\!\!\frac{k}{Ha}\!\right)^4
\!\Bigg[\!\Bigg(\!\frac{a^4}{6}\!-\!\frac{a^2}{2}\nonumber\\
&&\hspace{-1.2cm}-\frac{2a}{3}\!\Bigg)\!\ln(a)\!-\!\frac{59a^4}{120}\!+\!a^3\!+\!\frac{5a^2}{12}\!-\!\frac{5a}{6}\!-\!\frac{1}{8}
\!+\!\frac{a^{-1}}{30}\Bigg]\!\!+\!i\!\left(\!\!\frac{k}{Ha}\!\right)^5
\!\Bigg[\!\Bigg(\!\frac{4a^5}{75}\!-\!\frac{2a^3}{9}\!-\!\frac{7a^2}{15}\!-\!\frac{a}{4}\!-\!\frac{7}{45}\!\Bigg)\!\ln(a)\nonumber\\
&&\hspace{-1.2cm}-\frac{2783a^5}{15750}\!+\!\frac{2a^4}{5}\!+\!\frac{26a^3}{135}\!-\!\frac{13a^2}{90}\!+\!\frac{a}{30}\!-\!\frac{8197}{27000}
\!-\!\frac{a^{-2}}{840}\Bigg]\!\!-\!\left(\!\!\frac{k}{Ha}\!\right)^6\!\Bigg[\!\Bigg(\!\frac{2a^6}{135}
\!-\!\frac{a^4}{12}\!-\!\frac{2a^3}{9}\!-\!\frac{a^2}{8}
\!-\!\frac{a}{15}\!\Bigg)\nonumber\\
&&\hspace{0.5cm}\times\!\ln(a)\!-\!\frac{503a^6}{9450}\!+\!\frac{2a^5}{15}+\frac{19a^4}{240}\!-\!\frac{a^3}{810}\!-\!\frac{a^2}{48}\!-\!\frac{27a}{200}
\!-\!\frac{11}{3240}\!+\!\frac{a^{-1}}{840}\Bigg]
\!\!+\!\mathcal{O}\!\left(\!\Big(\!\frac{k}{Ha}\!\Big)^{7}\right)\!\!\Bigg\}\; .\label{g16power}
\ee
Analytic expression for one loop correction~(\ref{g1asSUM}), i.e.,
\beeq
\mathrm{g}_1(\eta,k)\!=\!\sum_{s=1}^6\mathrm{g}_{1,s}(\eta,k)\; ,\nonumber
\eneq
is obtained by adding up Eqs.~(\ref{g1mid}), (\ref{g2mid}), (\ref{g3mid}), (\ref{g4mid}), (\ref{intsing5}) and (\ref{mathcalA})-(\ref{mathcalB}), (\ref{intsing6}) and (\ref{mathcalC})-(\ref{mathcalD}). To save space we leave the addition to the reader. To obtain $\mathrm{g}_1(\eta,k)$ in powers of $\frac{k}{Ha}$ we use Eqs.~(\ref{g11power}), (\ref{g12power}), (\ref{g13power}), (\ref{g14power}), (\ref{g15power}) and (\ref{g16power}) in Eq.~(\ref{g1asSUM}),
\be
&&\hspace{-0.4cm}\mathrm{g}_1(\eta, k)\!=\!\mathrm{g}_0(0, k)\frac{1}{3}\Bigg\{\!i\!\left(\!\!\frac{k}{Ha}\!\right)\!\!\Bigg[a\!-\!\frac{3}{2}\!+\!\frac{a^{-2}}{2}\Bigg]
\!\!+\!\left(\!\!\frac{k}{Ha}\!\right)^2
\!\frac{1}{2}\Bigg[\frac{a^2}{4}\!-\!\frac{3a}{2}\!+\!\frac{9}{4}\!-\!a^{-1}\Bigg]\nonumber\\
&&\hspace{-0.4cm}+i\!\left(\!\!\frac{k}{Ha}\!\right)^3\!\frac{1}{3}
\Bigg[\!\frac{167a^3}{180}\!-\!\frac{9a^2}{4}\!+\!\frac{15a}{4}\!+\!\ln^2(a)\!-\!\frac{7}{3}\ln(a)
\!-\!\frac{41}{18}\!-\!\frac{3a^{-2}}{20}\Bigg]\!\!-\!\left(\!\!\frac{k}{Ha}\!\right)^4\!\frac{1}{4}
\Bigg[\frac{233a^4}{240}\!-\!\frac{7a^3}{4}\nonumber\\
&&\hspace{-0.4cm}+\frac{19a^2}{8}-\frac{43a}{12}\!+\!\frac{35}{16}\!-\!\frac{a^{-1}}{5}\Bigg]
\!\!-\!i\!\left(\!\!\frac{k}{Ha}\!\right)^5\!\!\frac{1}{12}
\Bigg[\!\frac{14113a^5}{10500}\!-\!\frac{9a^4}{4}\!+\!\frac{119a^3}{45}\!-\!\frac{4a^2}{5}\ln(a)
\!-\!\frac{107a^2}{30}\!+\!\frac{57a}{20}\nonumber\\
&&\hspace{-0.4cm}+\frac{2}{5}\!\ln^2(a)\!-\!\frac{46}{75}\ln(a)\!-\!\frac{2251}{2250}\!-\!\frac{3a^{-2}}{140}\Bigg]
\!\!+\!\left(\!\!\frac{k}{Ha}\!\right)^6\!\frac{1}{30}\Bigg[\frac{7681a^6}{6048}\!-\!\frac{31a^5}{16}
\!+\!\frac{77a^4}{64}\!-\!\frac{10a^3}{9}\ln(a)\!-\!\frac{17a^3}{16}\nonumber\\
&&\hspace{6.5cm}+\frac{65a^2}{32}\!-\!\frac{41a}{16}
\!+\!\frac{1921}{1728}\!-\!\frac{3a^{-1}}{56}\Bigg]
\!\!+\!\mathcal{O}\!\left(\!\Big(\!\frac{k}{Ha}\!\Big)^{7}\right)\!\!\Bigg\}\; .\label{g1pow}
\ee
This is the one-loop mode function for the Yukawa scalar. In the MMC $\lambda\varphi^4$ model, when the one-loop mode function $u_1(\eta, k)$ is expressed in $g_0(0, k)$ parenthesis, we get~\cite{VKO1} $k$-dependent terms as in Eq.~(\ref{Yukpwrspctrm}). We, however, also get $k$-independent but time dependent $\ln^2(a)$ and $\ln(a)$ corrections and a $k$-independent constant shift that is absorbed into a field strength renormalization. (See also Eq.~(\ref{power1}) in this paper.) In $\mathrm{g}_1(\eta, k)$, there is no $k$-independent shift which is also time independent to be absorbed into a field strength renormalization. Note also that $\mathrm{g}_1(\eta, k)$ approaches a constant for each mode $k$ in the late time limit. We study how the one-loop Yukawa scalar mode function tilts the scalar's power spectrum during inflation, in the next section.

\section{quantum-corrected Yukawa scalar power spectrum}
\label{sec:Power}

We computed one- and two-loop corrected power spectrum and two-point correlation function---at two distinct spacetime events---of the scalar field in MMC $\lambda\varphi^4$ model during de Sitter inflation in Refs.~\cite{VKO1} and~\cite{VKO2}, respectively. We worked out the effect of such a spectator field on the curvature power spectrum $\Delta^2_{\mathcal{R}}(k,t)$ in Ref.~\cite{zeta}. In various field theories, quantum effects on inflationary observables have been studied applying quantum~\cite{quant} and stochastic~\cite{stoch} formalisms. In this section, we compute the one-loop corrected power spectrum of a Yukawa scalar during inflation applying quantum field theory.

The power spectrum $\mathcal{P}(t, k)$ is defined as the Fourier transform
of the equal-time two-point correlation function of~the field, \be
\mathcal{P}_\phi(t, k)\!\equiv\!\!\int\!\! d^3x\,
e^{-i\vec{k}\cdot\vec{x}}\langle\Omega|\phi(t, \vec{x})
\phi(t, 0)|\Omega\rangle\; . \label{powerprim}\ee Hence, the two-point correlation function is the inverse Fourier transform of the power spectrum.

An alternative measure of power $\Delta^2_\phi (t, k)$ in a mode $k$ is obtained by considering the excess power in a bin of size $dk$ centered at $k$, $\frac{d^3k}{(2\pi)^3}\mathcal{P}_\phi(t, k)$. Integrating it
over all orientations of $\vec{k}$, we define\beeq
\frac{k^2dk}{2\pi^2}\mathcal{P}_\phi(t, k) \!\equiv\!
\frac{dk}{k}\Delta^2_\phi (t, k) \; .\label{Deltaphi}\eneq
Employing the full-field expansion
\beeq
\phi(t, \vec{x})=\!\int\!\!\frac{d^{D-1}k}{(2\pi)^{D-1}}
\Bigl\{\mathrm{g}(t, k) e^{i\vec{k} \cdot \vec{x}}\alpha(\vec{k})\!+\!
\mathrm{g}^*(t, k) e^{-i\vec{k} \cdot \vec{x}} \a^{\dagger}(\vec{k})
\Bigr\} \; , \label{freefieldexp}\eneq where $\mathrm{g}(t, k)$ is the amplitude of the quantum corrected mode function~[Eq.~(\ref{solpert})], in
Eqs.~(\ref{powerprim}) and (\ref{Deltaphi}) yields~\cite{MiaoPark} the quantum corrected power spectrum
\be
\mathcal{P}_{\phi}(t,
k)\!\!&=&\!\!|\mathrm{g}(t, k)|^2\\
\Delta^2_{\phi}(t, k)\!\!&=&\!\!\frac{k^3}{2\pi^2}\mathcal{P}_{\phi}(t,
k)\!=\!\frac{k^3}{2\pi^2}|\mathrm{g}(t, k)|^2 \; , \label{PowerDELTA}\ee
where the amplitude squared for each
mode\be \hspace{0cm}|\mathrm{g}(t, k)|^2 \!=\!|\mathrm{g}_0(t,
k)|^2\!+\!\frac{f^2}{4\pi^2} \Big[\mathrm{g}_0(t, k)\mathrm{g}_1^*(t, k)\!+\!\mathrm{g}_0^*(t,
k)\mathrm{g}_1(t, k)\Big]\!+\!\mathcal{O}\!\left(\!\Big(\!\frac{f^2}{4\pi^2}\!\Big)^2\right) \;
.\label{Phisquared}\ee
Note that the tree-order power spectrum\beeq\Delta^2_{{\phi_0}}(t,
k)\!=\!\frac{k^3}{2\pi^2}|\mathrm{g}_0(t, k)|^2\!=\!\frac{H^2}{4\pi^2}\!\left[\!1\!+\!\frac{k^2}{H^2}a^{-2}\right] \; , \eneq
is the scale-invariant Harrison-Zel'dovich spectrum, in the late time limit.

Employing Eq.~(\ref{g0}) and the analytic expression for $g_1$---obtained by using Eqs.~(\ref{g1mid}), (\ref{g2mid}), (\ref{g3mid}), (\ref{g4mid}), (\ref{intsing5}), (\ref{mathcalA}), (\ref{mathcalB}), (\ref{intsing6}),  (\ref{mathcalC}), (\ref{mathcalD}) in Eq.~(\ref{g1asSUM})---in Eqs.~(\ref{PowerDELTA}) and (\ref{Phisquared}) one gets an analytic result for the quantum corrected power spectrum at one-loop order. We express it as a series in powers of $\frac{k}{Ha}$
\be
&&\hspace{-0.4cm}\Delta^2_\phi(t,
k)\!=\!\frac{H^2}{4\pi^2}\Bigg\{\!1\!+\!\left(\!\!\frac{k}{Ha}\!\right)^2\!\!\!+\!\frac{f^2}{4\pi^2}\frac{1}{9} \Bigg\{\!\!\left(\!\!\frac{k}{Ha}\!\right)^2
\!\Bigg[\!\frac{3a^2}{4}\!-\!\frac{9a}{2}\!+\!\frac{27}{4}\!-\!3a^{-1}\!\Bigg]
\!\!-\!\left(\!\!\frac{k}{Ha}\!\right)^4\!\Bigg[\!\frac{233a^4}{160}\!-\!\frac{21a^3}{8}\nonumber\\
&&\hspace{-0.4cm}+\frac{51a^2}{16}-\frac{41a}{8}\!+\!\frac{93}{32}
\!+\!\frac{6a^{-1}}{5}\!-\!a^{-2}\Bigg]\!\!+\!\left(\!\!\frac{k}{Ha}\!\right)^6
\!\Bigg[\!\frac{7681a^6}{30240}\!-\!\frac{31a^5}{80}\!-\!\frac{39a^4}{80}\!-\!\frac{2a^3}{9}\ln(a)
\!+\!\frac{232a^3}{135}\!-\!\frac{95a^2}{32}\nonumber\\
&&\hspace{-0.4cm}+\frac{403a}{80}\!+\!\frac{2}{3}\ln^2(a)\!-\!\frac{14}{9}\ln(a)\!-\!\frac{1253}{360}
\!+\!\frac{18a^{-1}}{35}\!-\!\frac{a^{-2}}{5}\!\Bigg]\!\!-\!\left(\!\!\frac{k}{Ha}\!\right)^8\!\!\frac{1}{30}
\Bigg[\!\frac{33043a^8}{53760}\!-\!\frac{381a^7}{448}\!-\!\frac{37445a^6}{8064}\nonumber\\
&&\hspace{-0.4cm}-\frac{4a^5}{5}\!\ln(a)\!+\!\frac{527279a^5}{33600}\!-\!\frac{5727a^4}{256}\!+\!\frac{8a^3}{3}\!\ln(a)
\!+\!\frac{16075a^3}{576}\!-\!4a^2\!\ln(a)\!-\!\frac{15905a^2}{384}\!+\!\frac{22299a}{448}\!+\!4\!\ln^2(a)\nonumber\\
&&\hspace{2cm}-\frac{116}{15}\ln(a)\!-\!\frac{2338219}{89600}
\!+\!\frac{40a^{-1}}{21}\!-\!\frac{18a^{-2}}{35}\!\Bigg]
\!\!+\!\mathcal{O}\!\left(\!\Big(\!\frac{k}{Ha}\!\Big)^{10}\right)\!\!\Bigg\}
\!+\!\mathcal{O}\!\left(\!\Big(\!\frac{f^2}{4\pi^2}\!\Big)^2\right)\!\Bigg\}\; .\label{Yukpwrspctrm}
\ee
Unlike the MMC $\lambda\varphi^4$ model, where we get~\cite{VKO1} $k$-independent but time dependent $\ln^2(a)$ and $\ln(a)$ corrections in the one-loop power spectrum (see also Eq.~(\ref{powervarphi}) in this paper), there is no $k$-independent correction in Eq.~(\ref{Yukpwrspctrm}). The $\Delta^2_\phi(t, k)$ approaches a constant for each mode $k$ in the late time limit.

The spectral index $n$ measures the variation of the power
spectrum of fluctuations in a field with scale. We define
the spectral index $n_\phi$ of our Yukawa coupled scalar
the same way as for the graviton, \beeq n_\phi(t,
k)\!\equiv\!\frac{d\ln\left(\Delta^2_\phi(t, k)\right)}{d\ln(k)}\!=\!\frac{1}{\Delta^2_\phi(t,
k)}\!\left[k\frac{\partial}{\partial
k}\!+\!a(t)\frac{\partial}{\partial
a(t)}\right]\!\Delta^2_\phi(t, k) \;
.\label{spectral}\eneq
During inflation many modes exit the horizon. This ``first horizon crossing'' happens for a mode with physical wave number $k_{\rm phys}(t)\!=\!k/a(t)$ at comoving time $t_k$ when
$k_{\rm phys}(t_k)\!=\!H(t_k)$. During de Sitter inflation,  $H$ is
constant so this mode has comoving wave number\beeq k\!=\!H a(t_k)\!=\!He^{Ht_k}\;
.\label{MODEcrossing}\eneq The spectral index $n_\phi$ for the mode $k$ at $t_k$ is computed using
Eqs.~(\ref{Yukpwrspctrm})-(\ref{MODEcrossing}). Up to $\mathcal{O}((\!\frac{f^2}{4\pi^2})^2)$, we find\be &&\hspace{-0.4cm}
n_\phi(t_k, k)\!=\!\frac{f^2}{4\pi^2}\frac{1}{6}\Bigg\{\!\!\!\left(\!\!\frac{k}{Ha}\!\right)^2
\!\!\Bigg[a^2\!-\!3a\!+\!2a^{-1}\Bigg]\!\!-\!\left(\!\!\frac{k}{Ha}\!\right)^4\!\!\Bigg[\frac{233a^4}{60}
\!-\!\frac{21a^3}{4}\!+\!\frac{21a^2}{4}
\!-\!\frac{77a}{12}\!+\!\frac{6a^{-1}}{5}\!+\!\frac{4a^{-2}}{3}\Bigg]\nonumber\\
&&\hspace{-0.4cm}+\!\left(\!\!\frac{k}{Ha}\!\right)^6\!\!\Bigg[\!\frac{7681a^6}{7560}\!-\!\frac{31a^5}{24}
\!+\!\frac{31a^4}{12}\!-\!\frac{4a^3}{9}\!\ln(a)\!-\!\frac{353a^3}{180}\!+\!\frac{31a^2}{24}
\!-\!\frac{367a}{120}\!+\!\frac{8}{9}\!\ln(a)\!-\!\frac{28}{27}
\!+\!\frac{6a^{-1}}{7}\!+\!\frac{8a^{-2}}{5}\Bigg]\nonumber\\
&&\hspace{-0.4cm}-\!\left(\!\!\frac{k}{Ha}\!\right)^8
\!\frac{1}{9}\Bigg[\!\frac{33043a^8}{33600}\!-\!\frac{381a^7}{320}\!+\!\frac{1143a^6}{320}\!-\!\frac{4a^5}{5}\!\ln(a)
\!+\!\frac{131303a^5}{33600}\!+\!\frac{1713a^4}{320}
\!-\!\frac{12a^3}{5}\!\ln(a)\!-\!\frac{119a^3}{320}\nonumber\\
&&\hspace{-0.4cm}-\frac{8a^2}{5}\!\ln(a)
\!-\!\frac{5513a^2}{960}\!-\!\frac{39357a}{2240}\!+\!\frac{48}{5}\ln(a)\!-\!\frac{272}{25}
\!+\!\frac{22a^{-1}}{3}\!+\!\frac{2556a^{-2}}{175}\Bigg]
\!\!+\!\mathcal{O}\!\left(\!\Big(\!\frac{k}{Ha}\!\Big)^{10}\right)\!\!\Bigg\}\; . \label{indexYukawa} \ee The $a$-dependent terms in Eq.~(\ref{indexYukawa}) are to be evaluated
at $t\!=\!t_k$. Unlike the MMC $\lambda\varphi^4$ model where we get~\cite{VKO1} $k$-independent but time dependent $\ln(a)$ correction in the one-loop spectral index (see also Eq.~(\ref{spectrvarphi}) in this paper), there isn't any $k$-independent correction in Eq.~(\ref{Yukpwrspctrm}). And, for each mode $k$, the index $n_\phi(t, k)$ asymptotes to a constant in the late time limit.

The running of the spectral
index\beeq \alpha_\phi(t, k)\!\equiv\!\frac{d\,n_\phi(t,
k)}{d\ln(k)}\!=\!\left[k\frac{\partial}{\partial
k}\!+\!a(t)\frac{\partial}{\partial
a(t)}\right]\!n_\phi(t, k)\; ,\label{running}\eneq measures how the spectral index
changes as the scale varies. At $t\!=\!t_k$, we have \be &&\hspace{-0.3cm}\alpha_\phi(t_k,
k)\!=\!\frac{f^2}{4\pi^2}\frac{1}{3}\Bigg\{\!\!\left(\!\!\frac{k}{Ha}\!\right)^2
\!\Bigg[a^2\!\!-\!\frac{3a}{2}\!-\!a^{-1}\!\Bigg]
\!\!-\!\left(\!\!\frac{k}{Ha}\!\right)^4\!\Bigg[\!\frac{233a^4}{30}\!-\!\frac{63a^3}{8}\!+\!\frac{21a^2}{4}
\!-\!\frac{77a}{24}\!-\!\frac{3a^{-1}}{5}\!-\!\frac{4a^{-2}}{3}\!\Bigg]\nonumber\\
&&\hspace{-0.3cm}+\!\left(\!\!\frac{k}{Ha}\!\right)^6
\!\Bigg[\!\frac{7681a^6}{2520}\!-\!\frac{155a^5}{48}\!+\!\frac{31a^4}{6}
\!-\!\frac{2a^3}{3}\!\ln(a)\!-\!\frac{1139a^3}{360}\!+\!\frac{31a^2}{24}
\!-\!\frac{367a}{240}\!+\!\frac{4}{9}
\!-\!\frac{3a^{-1}}{7}\!-\!\frac{8a^{-2}}{5}\!\Bigg]\nonumber\\
&&\hspace{-0.3cm}-\!\left(\!\!\frac{k}{Ha}\!\right)^8\!\frac{1}{3}
\Bigg[\!\frac{33043a^8}{25200}\!-\!\frac{889a^7}{640}\!+\!\frac{1143a^6}{320}\!-\!\frac{2a^5}{3}\!\ln(a)
\!+\!\frac{125927a^5}{40320}\!+\!\frac{571a^4}{160}
\!-\!\frac{6a^3}{5}\!\ln(a)\!-\!\frac{75a^3}{128}\nonumber\\
&&\hspace{-0.3cm}-\frac{8a^2}{15}\!\ln(a)\!-\!\frac{6281a^2}{2880}
\!-\!\frac{13119a}{4480}\!+\!\frac{8}{5}
\!-\!\frac{11a^{-1}}{9}\!-\!\frac{852a^{-2}}{175}\!\Bigg]
\!+\!\mathcal{O}\!\left(\!\Big(\!\frac{k}{Ha}\!\Big)^{10}\right)\!\!\Bigg\}
\!+\!\mathcal{O}\!\left(\!\Big(\!\frac{f^2}{4\pi^2}\!\Big)^2\right)\; .\label{runnYukawa}
\ee
Like the one-loop running in MMC $\lambda\varphi^4$ model~\cite{VKO1} (see also Eq.~(\ref{runnvarphi}) in this paper), one loop running~(\ref{runnYukawa}) of the Yukawa scalar
approaches a constant for each mode $k$ in the late time limit.

We infer from
Eqs.~(\ref{Yukpwrspctrm}), (\ref{indexYukawa}) and (\ref{runnYukawa}) that the one-loop effect
blue tilts the spectrum. Therefore, the amplitudes of fluctuations grow slightly
towards the smaller scales. This quantum effect is in contrast with that of the MMC $\lambda\varphi^4$ model where the spectrum is red tilted~\cite{VKO1} and the amplitudes grow toward the larger scales.

Our loop corrections~(\ref{Yukpwrspctrm}), (\ref{indexYukawa}) and (\ref{runnYukawa}) are time-dependent. To make a reasonable estimate of the corrections
we define the number of e-foldings $N$ after first horizon crossing until the end of inflation. See the discussion in Ref.~\cite{VKO1}. For the window of observable modes, typical value for $N\!\sim\!50$.
Hence we may take $a\!\sim\!e^{50}$. This implies that our one-loop corrections become almost constants for the observable modes.

Next, we apply the computation method used for the Yukawa scalar in this paper, to obtain exact analytical expressions---instead of series expansions~\cite{VKO1}---for the cosmological entities $\Delta^2_\varphi$, $n_\varphi$ and $\alpha_\varphi$ in the MMC $\lambda\varphi^4$ model during inflation.

\section{Comparison with the massless minimally coupled $\lambda\varphi^4$ model}
\label{sec:MMCS}

Let us consider a MMC scalar $\varphi(x)$ with a quartic self-interaction in the locally de Sitter background. The quantum corrected scalar mode function $\varphi(x;\vec{k})$ is
the solution of the linearized effective field
equation \beeq \dd_\mu\Bigl(\sqrt{-g}
g^{\mu\nu} \partial_{\nu} \varphi(x)\Bigr) \!=\!\!
\int_{\eta_i}^0 \!\!\!d\eta' \!\!\int\!\!
d^3x'M^2(x;x') \varphi(x') \; ,
\label{lineqn} \eneq where $M^2(x;x')$ is the self-mass squared of the scalar. We solve the effective field equation
perturbatively by expanding $\varphi(x;\vec{k})$ and $M^2(x;x')$
in powers of $\lambda$ as
\be\varphi(x;\vec{k})\!\!&\equiv&\!\!u(\eta; k)
e^{i \vec{k} \cdot\vec{x}}\!\equiv\!\sum_{\ell=0}^{\infty}
\l^{\ell} u_{\ell}(\eta,k) e^{i \vec{k} \cdot\vec{x}}\;
,\label{pertmode}\\
M^2(x;x')\!\!&\equiv&\!\!\sum_{\ell=0}^{\infty} \l^{\ell}
\mathcal{M}^2_{\ell}(x;x')\; .\label{pertmass}
\ee
Note that the tree-order ($\ell\!=\!0$) solution $u_0(\eta,k)$ is the
Bunch-Davies mode function $\mathrm{g}_0(\eta,k)$ given in Eq.~(\ref{g0}). Henceforth, we use $\mathrm{g}_0(\eta, k)$ instead of $u_0(\eta, k)$. The one-loop mass squared term~\cite{BOW}
\beeq \mathcal{M}^2_1(x;x')\!=\!\frac{H^2}{8\pi^2}a^4\ln(a) \delta^4(x\!-\!x')\; .\label{1loopmass}\eneq (The self-mass squared remains perturbatively small and does not go tachyonic~\cite{sta}.) Using this result and Eqs.~(\ref{pertmode})-(\ref{pertmass}) in Eq.~(\ref{lineqn}), we
obtain an integro-differential equation---the analog of Eq.~(\ref{Integrog1})---for the order $\lambda$
correction $u_1(\eta, \vec{k})$, \be \Bigl[
\frac{\partial^2}{\partial t^2} \!+\! 3 H \frac{\partial}{\partial
t} \!+\! \frac{k^2}{a^2}\Bigr] u_1 \!\!&=&\!\!
-\frac{1}{a^4}\!\!\int_{\eta_i}^0\!\!\! d\eta' \mathrm{g}_0(\eta',
k)\!\!\int\!\!d^3x' \mathcal{M}^2_1(x;x')e^{-i \vec{k}
\cdot(\vec{x}-\vec{x}\,')}\nonumber\\
\!\!&=&\!\!-\frac{H^2}{8\pi^2} \mathrm{g}_0(\eta,k)\ln(a)\; .
\label{IntegroPhi1}\ee The solution $u_1(\eta,k)$ of
Eq.~(\ref{IntegroPhi1}) can be written as an integral over
comoving time as \beeq
u_1(\eta,k)\!=\!-\frac{H^2}{8\pi^2}\!\!\int_0^t\!\! dt'
G(t,t';k)\, \mathrm{g}_0(\eta',k) \ln\left(a(\eta')\right)\; ,
\label{Phi1}\eneq where the Green's function $G(t,t';k)$ is given in Eq.~(\ref{Green}). Hence, we have \be
&&\hspace{-0.5cm}u_1(\eta,k)\!=\!-\frac{iH}{8\pi^2}\!\int_1^a\!\!\! da'
a'^2\ln(a')\Bigg[\mathrm{g}_0(\eta, k)|\mathrm{g}_0(\eta', k)|^2\!-\!\mathrm{g}_0^*(\eta, k)\mathrm{g}_0^2(\eta', k)\Bigg]\label{Phi1varphiara}\\
&&\hspace{1cm}\!\equiv\!-\frac{iH^3}{16\pi^2k^3}\Bigg\{\!\mathrm{g}_0(\eta, k)\mathcal{F}(\eta, k)\!-\!\mathrm{g}^*_0(\eta, k)\mathcal{G}(\eta, k)\!\Bigg\}\; ,\label{analyticalphi1}
\ee
where
\be
&&\hspace{-1cm}\mathcal{F}(\eta, k)\!\equiv\!\!\int_1^a\!\!\!
da'\!\left[a'^2\!+\!\frac{k^2}{H^2}\right]\!\ln(a')\!=\!\frac{a^3}{3}\!\left(\!\ln(a)\!-\!\frac{1}{3}\!\right)
\!+\!\frac{1}{9}
\!+\!\frac{k^2}{H^2}\!\left[a\Big(\!\ln(a)\!-\!1\!\Big)\!+\!\!1\right]\label{mathcalF}\; ,\\
&&\hspace{-1cm}\mathcal{G}(\eta, k)\!\equiv\!\!\int_1^a\!\!\!da'\!\left[a'\!-\!\frac{ik}{H}\right]^2\!\!\!e^{\frac{2ik}{Ha'}}\!\ln(a')\!=\!\frac{a}{3}e^{\frac{2ik}{Ha}}
\!\left\{\!a\!\left[\ln(a)\!-\!\frac{1}{3}\right]\!\!
\left[a\!-\!\frac{2ik}{H}\right]
\!\!+\!\!\frac{k^2}{H^2}\!\left[\ln(a)\!-\!\frac{7}{3}\right]\!\right\}\nonumber\\
&&\hspace{-1cm}+\frac{1}{9}e^{\frac{2ik}{H}}\!\left\{\!1\!-\!\frac{2ik}{H}\!+\!7\frac{k^2}{H^2}\!\right\}
\!-\!\frac{i}{3}\frac{k^3}{H^3}\Bigg\{\!\!\ln^2(a)\!+\!\!2\!\ln(a)\!\Bigg[{\rm ci}\Big(\!\frac{2k}{Ha}\Big)\!-\!\ln\!\Big(\!\frac{2k}{H}\Big)\!-\!\gamma\!+\!i\!\left(\!\frac{\pi}{2}\!+\!{\rm si}\Big(\!\frac{2k}{Ha}\Big)\!\right)\!\!\Bigg]\nonumber\\
&&\hspace{-1cm}-\frac{14}{3}\!\Bigg[{\rm ci}\Big(\!\frac{2k}{Ha}\Big)\!-\!{\rm ci}\Big(\!\frac{2k}{H}\Big)\!+\!i\!\left(\!{\rm si}\Big(\!\frac{2k}{Ha}\Big)\!-\!{\rm si}\Big(\!\frac{2k}{H}\Big)\!\right)\!\!\Bigg]\!\Bigg\}\!+\!\frac{4}{3}\frac{k^4}{H^4}\Bigg\{\!a^{-1} {}_{3}\mathcal{F}_{3}\!\left(\!\!1, 1, 1; 2, 2, 2; \frac{2ik}{Ha}\!\right)\nonumber\\
&&\hspace{9.2cm}-{}_{3}\mathcal{F}_{3}\!\left(\!\!1, 1, 1; 2, 2, 2; \frac{2ik}{H}\!\right)\!\!\Bigg\}\; .\label{mathcalG}
\ee We employed integrals~(\ref{intg11int1}) and (\ref{intg11int2}) in Eqs.~(\ref{mathcalF}) and (\ref{mathcalG}).
Analytical exact result~(\ref{analyticalphi1}) agrees with the series solution\be
&&\hspace{-0.5cm}u_1(\eta,k)\!=\!-\frac{iH^3}{16\pi^2k^3}\Bigg\{\!\mathrm{g}_0(\eta,
k)\!\!\left[a^3\frac{\ln(a)}{3}\!-\!\frac{a^3}{9}\!+\!\frac{1}{9}\!+\!\frac{k^2}{H^2}\Big(a\ln(a)
\!-\!a\!+\!1\Big)\right]\!\!-\!\mathrm{g}_0^*(\eta,
k)\!\Bigg[\frac{ik^3}{H^3}\frac{\ln^2(a)}{3}\nonumber\\
&&\hspace{1.35cm}+\!\sum_{n \doteq\,
0}^\infty\!\frac{1}{n!}\!\left(\frac{2ik}{H}\right)^n\!\!
\Bigg(\!\!\left[\frac{a^{3-n}}{3\!-\!n}\ln(a)\!-\!\frac{(a^{3-n}\!-\!1)}{(3\!-\!n)^2}\right]
\!\!-\!\frac{2ik}{H}\!
\left[\frac{a^{2-n}}{2\!-\!n}\ln(a)\!-\!\frac{(a^{2-n}\!-\!1)}{(2\!-\!n)^2}\right]\nonumber\\
&&\hspace{8cm}-\frac{k^2}{H^2}\!\left[\frac{a^{1-n}}{1\!-\!n}\ln(a)
\!-\!\frac{(a^{1-n}\!-\!1)}{(1\!-\!n)^2}\right]\!\!\Bigg)\!\Bigg]\!\Bigg\}\;
, \label{exactPhi1} \ee that we obtained in Ref.~\cite{VKO1}. (Definition of the symbol $\doteq$ in the sum is given after Eq.~(\ref{exactg11asasum}), in this paper.) When it is expressed in powers of $\frac{k}{Ha}$, we have \be&&\hspace{-0.35cm}u_1(\eta,k)\!=\!\frac{\mathrm{g}_0(0,
k)}{2^4\,3\,\pi^2}\Bigg\{\!\!\!-\!\ln^2(a)\!+\!\frac{2\ln(a)}{3}
\!-\!\frac{2}{9}\!+\!\frac{2a^{-3}}{9}
\!-\!\!\left(\!\!\frac{k}{Ha}\!\right)^2\!\Bigg[\frac{a^2}{5}
\!+\!\frac{\ln^2(a)}{2}\!-\!\frac{7\ln(a)}{3}
\!+\!\frac{16}{9}\!-\!\frac{2}{a}\!+\!\frac{a^{-3}}{45}\Bigg]\nonumber\\
&&\hspace{-0.35cm}-i\!\left(\!\!\frac{k}{Ha}\!\right)^3\!\Bigg[\frac{2a^3}{27}\!-\!\frac{\ln^2(a)}{3}
\!-\!\frac{2\ln(a)}{9}
\!-\!\frac{2}{27}\Bigg]\!\!+\!\!\left(\!\!\frac{k}{Ha}\!\right)^4\!\Bigg[\frac{3a^4}{140}\!-\!\frac{a^2}{10}
\!+\!\frac{\ln^2(a)}{8}\!-\!
\frac{\ln(a)}{12}\!+\!\frac{5}{18}\!-\!\frac{a^{-1}}{5}
\!+\!\frac{a^{-3}}{1260}\Bigg]\nonumber\\
&&\hspace{2.5cm}+i\!\left(\!\!\frac{k}{Ha}\!\right)^5\!\Bigg[\frac{2a^5}{375}\!-\!\frac{a^3}{27}
\!+\!\frac{a^2}{15}\!-\!\frac{\ln^2(a)}{30}
\!-\!\frac{11\ln(a)}{225}\!-\!\frac{118}{3375}\Bigg]\!\!+\!\mathcal{O}\!\left(\!\Big(\!\frac{k}{Ha}\!\Big)^{6}\right)\!\Bigg\} \;
.\label{power1}\ee The
$k$-independent term $-2/9$ in Eq.~(\ref{power1}), which is also
time independent, can be absorbed~\cite{VKO1} into a field strength
renormalization. Hence it is not an
observable. We can shift the one-loop mode function $u_1(\eta,k)$ so that the term \beeq
-\frac{\mathrm{g}_0(0,
k)}{2^3\,3^3\,\pi^2}\; ,\label{cnstntshift}\eneq is removed.

Now, to relate the case to the Yukawa coupled scalar model, let us consider the $\mathrm{g}_{1,1}$ term given in Eq.~(\ref{g11intsilk}). It can be recast as
\be
&&\hspace{-0.4cm}\mathrm{g}_{1,1}(\eta,k)\!=\!-iH\Bigg\{\!\Bigg[\mathrm{g}_0(\eta, k)\!\!\!\int_1^a\!\!\! da'
a'^2\ln(a')|\mathrm{g}_0(\eta', k)|^2\!-\!\mathrm{g}_0^*(\eta, k)\!\!\!\int_1^a\!\!\! da'
a'^2\ln(a')\mathrm{g}_0^2(\eta', k)\Bigg]\!\!+\!\!\Bigg[{\rm C.C.}\Bigg]\!\Bigg\}\nonumber \; .
\ee
Comparing this result with Eq.~(\ref{Phi1varphiara}) we can express the $\mathrm{g}_{1,1}$ in terms of the one-loop mode function $u_1(\eta, k)$ in the MMC $\lambda\varphi^4$ model,
\be
\mathrm{g}_{1,1}(\eta,k)\!=\!8\pi^2\left[u_1(\eta, k)\!-\!u^*_1(\eta, k)\right]\; .
\ee

The full field expansion
\beeq
\varphi(x)\!=\!\!\int\!\!\frac{d^{D-1}k}{(2\pi)^{D-1}}
\Bigl\{u(t, k) e^{i\vec{k} \cdot \vec{x}}\alpha(\vec{k})\!+\!
u^*(t, k) e^{-i\vec{k} \cdot \vec{x}} \a^{\dagger}(\vec{k})
\!\Bigr\} \; , \eneq yields
\beeq
\Delta^2_\varphi(t,
k)\!=\!\frac{k^3}{2\pi^2} |u(t, k)|^2\; ,
\eneq
where
\beeq
|u(t, k)|^2\!=\!|\mathrm{g}_0(t, k)|^2\!+\!\lambda\Big[\mathrm{g}^*_0(t, k)u_1(t, k)\!+\!\mathrm{g}_0(t, k)u^*_1(t, k)\Big]\!\!+\!\mathcal{O}(\lambda^2)\; .\label{Phiabsolutesqrd}
\eneq
If we naively insert $u_1(t, k)$ given in Eq.~(\ref{analyticalphi1}) into Eq.~(\ref{Phiabsolutesqrd}), we find
\be
|u(t, k)|^2\!=\!\frac{H^2}{2k^3}\Bigg\{\!1\!+\!\!\left(\!\!\frac{k}{Ha}\!\right)^2
\!\!\!+\!\frac{\lambda H}{8\pi^2}i\Bigg[{\mathrm{g}_0^*}^2(\eta, k)\mathcal{G}(\eta, k)\!-\!\mathrm{g}_0^2(\eta, k)\mathcal{G}^*(\eta, k)\Bigg]\!\!+\!\mathcal{O}(\lambda^2)\!\Bigg\}\;
,\label{quantcorrampsquared}
\ee
and obtain an analytical expression for the quantum corrected power spectrum
\be
\Delta^2_\varphi(t,
k)\!=\!\frac{H^2}{4\pi^2}\Bigg\{\!1\!+\!\!\left(\!\!\frac{k}{Ha}\!\right)^2\!\!\!-\!\frac{\lambda }{8\pi^2}\!\Bigg[\mathcal{N}(t, k)\!+\!\mathcal{N}^*(t, k)\Bigg]\!\!+\!\mathcal{O}(\lambda^2)\!\Bigg\}\; ,\label{powerwithN}
\ee
where
\be
&&\hspace{-0.3cm}\mathcal{N}(t, k)\!=\!\frac{2}{3}\ln(a)\!-\!\frac{8}{9}
+\!i\!\left[\frac{H}{k}\!-\!\frac{i}{a}\right]^2\!\!\!e^{\frac{2ik}{Ha}}
\Bigg\{\!\frac{e^{-\frac{2ik}{H}}}{18}\!\left[\frac{H}{k}\!+\!2i\!+\!7\frac{k}{H}\right]
\!\!-\!i\frac{k^2}{H^2}\Bigg\{\!\frac{\ln(a)}{3}\Bigg[\!\frac{\ln(a)}{2}\!+\!\gamma\!+\!\ln\!\Big(\!\frac{2k}{Ha}\!\Big)\nonumber\\
&&\hspace{-0.3cm}-{\rm ci}\Big(\!\frac{2k}{Ha}\!\Big)\!+\!i\!\left\{\!\frac{\pi}{2}\!+\!{\rm si}\Big(\!\frac{2k}{Ha}\!\Big)\!\right\}\!\!\Bigg]\!\!+\!\frac{7}{9}\Bigg[{\rm ci}\Big(\!\frac{2k}{Ha}\!\Big)\!\!-\!{\rm ci}\Big(\!\frac{2k}{H}\!\Big)\!\!-\!i\!\left\{\!{\rm si}\Big(\!\frac{2k}{Ha}\!\Big)\!\!-\!{\rm si}\Big(\!\frac{2k}{H}\!\Big)\!\right\}\!\!\Bigg]\!\Bigg\}\nonumber\\
&&\hspace{2.5cm}+\frac{k^3}{H^3}\frac{2}{3}\Bigg[a^{-1}{}_{3}\mathcal{F}_{3}\!\left(\!\!1, 1, 1; 2, 2, 2; -\frac{2ik}{Ha}\!\right)\!-\!{}_{3}\mathcal{F}_{3}\!\left(\!\!1, 1, 1; 2, 2, 2; -\frac{2ik}{H}\!\right)\!\!\Bigg]\!\Bigg\}\; ,
\ee
which is $iH\mathrm{g}_0^2\mathcal{G}^*$ up to an additional pure imaginary part that will be canceled out when $\mathcal{N}$ is added up with $\mathcal{N}^*$ in Eq~(\ref{powerwithN}).  Note that using the shifted one-loop mode function $u_1(t, k)$ (see Eqs.~(\ref{power1})-(\ref{cnstntshift})) in Eq.~(\ref{Phiabsolutesqrd}) yields \beeq
|u(t, k)|^2\rightarrow|u(t, k)|^2\!+\!\frac{H^2}{2k^3}\frac{\lambda}{2^2 3^3 \pi^2}\!\left[\cos\!\Big(\!\frac{k}{Ha}\!\Big)\!+\!\frac{k}{Ha}\!\sin\!\Big(\!\frac{k}{Ha}\!\Big)\right]\; ,
\eneq
which induces a time dependent shift \beeq
\mathcal{N}(t, k)\rightarrow \mathcal{N}(t, k)\!-\!\frac{1}{3^3}\!\left[\cos\!\Big(\!\frac{k}{Ha}\!\Big)
\!+\!\frac{k}{Ha}\!\sin\!\Big(\!\frac{k}{Ha}\!\Big)\right]\; ,\label{additinN}
\eneq
in Eq.~(\ref{powerwithN}) and therefore alters the power spectrum. When expressed in powers of $\frac{k}{Ha}$ we obtain
\be
&&\hspace{-0.4cm}\Delta^2_\varphi(t,
k)\!=\!\frac{H^2}{4\pi^2}\Bigg\{\!\!1\!+\!\frac{k^2}{H^2}a^{-2}\!-\!\frac{\lambda}{2^3\,3\,\pi^2}\Bigg[\!\ln^2(a)
\!-\!\frac{2\ln(a)}{3}\!-\!\frac{2a^{-3}}{9}\!+\!\left(\!\!\frac{k}{Ha}\!\right)^2\!\!\!\Big(\!\frac{a^2}{5}
\!+\!\ln^2(a)
\!-\!\frac{8\ln(a)}{3}\!+\!\frac{16}{9}\nonumber\\
&&\hspace{-0.4cm}-2a^{-1}\!-\!\frac{4a^{-3}}{45}\!\Big)
\!-\!\left(\!\!\frac{k}{Ha}\!\right)^4\!\!\Big(\!\frac{3a^4}{140}\!-\!\frac{a^{2}}{5}\!+\!\ln(a)
\!-\!\frac{11}{18}\!+\!\frac{4a^{-1}}{5}\!-\!\frac{4a^{-3}}{105}\!\Big)
\!+\!\left(\!\!\frac{k}{Ha}\!\right)^6\!\!\Big(\!\frac{2a^6}{1701}\!-\!\frac{3a^{4}}{140}\!+\!\frac{4a^{3}}{81}
\nonumber\\
&&\hspace{2.5cm}-\frac{2\ln^2(a)}{9}\!+\!\frac{7\ln(a)}{27}\!-\!\frac{239}{648}
\!+\!\frac{12a^{-1}}{35}\!-\!\frac{8a^{-3}}{1701}\!\Big)\!+\!\mathcal{O}\!\left(\!\Big(\!\frac{k}{Ha}\!\Big)^{8}
\right)\!\Bigg]\!\!+\!\mathcal{O}(\lambda^2)\!\Bigg\}\; .\label{powervarphi}
\ee
Because the logarithmic derivative of induced shift in $\mathcal{N}$,
\beeq
\left[k\frac{\partial}{\partial
k}\!+\!a\frac{\partial}{\partial
a}\right]\!\left\{\!-\frac{1}{3^3}\!\left[\cos\!\Big(\!\frac{k}{Ha}\!\Big)
\!+\!\frac{k}{Ha}\!\sin\!\Big(\!\frac{k}{Ha}\!\Big)\right]\!\right\}\!=\!0\; ,
\eneq
it has no effect at all in the numerator of the spectral index. We find\be &&\hspace{-1cm}n_\varphi(t_k, k)\!=\!\frac{1}{\Delta^2_\varphi(t_k,
k)}\!\Bigg\{\!\!\left(\!-\frac{\lambda H^2}{48\pi^4}\!\right)\!\Bigg\{\!2\!+\!\frac{H^2}{k^2}\!\left[\!1\!-\!a^{-1}\!\!-\!(a\!+\!1)
\Big(\!\frac{k}{Ha}\!\Big)^2\right]\!\cos\!\Big(\frac{2k}{H}\!\left(1\!-\!a^{-1}\right)\!\Big)\nonumber\\
&&\hspace{-1cm}-\frac{H}{2k}\!\left[\!
1\!+\!\frac{H^2}{k^2}\!+\!4a^{-1}\!\!-\!a^{-2}\!\!-\!\Big(\!\frac{k}{Ha}\!\Big)^2\right]
\!\sin\!\Big(\frac{2k}{H}\!\left(1\!-\!a^{-1}\right)\!\Big)
\!+\!\Bigg\{\!\!\left[\!1\!-\!\Big(\!\frac{k}{Ha}\!\Big)^2\right]\!\!\left[{\rm ci}\Big(\!\frac{2k}{H}\!\Big)\!\!-\!{\rm ci}\Big(\!\frac{2k}{Ha}\!\Big)\right]\nonumber\\
&&\hspace{-1cm}-\frac{2k}{Ha}\!\left[{\rm si}\Big(\!\frac{2k}{H}\!\Big)\!\!-\!{\rm si}\Big(\!\frac{2k}{Ha}\!\Big)\right]\!\!\Bigg\}
\cos\!\Big(\!\frac{2k}{Ha}\!\Big)\!+\!\Bigg\{\!\!\left[\!1\!-\!\Big(\!\frac{k}{Ha}\!\Big)^2\right]\!\!\left[{\rm si}\Big(\!\frac{2k}{H}\!\Big)\!\!-\!{\rm si}\Big(\!\frac{2k}{Ha}\!\Big)\right]\nonumber\\
&&\hspace{5.2cm}+\frac{2k}{Ha}\!\left[{\rm ci}\Big(\!\frac{2k}{H}\!\Big)\!\!-\!{\rm ci}\Big(\!\frac{2k}{Ha}\!\Big)\right]\!\!\Bigg\}
\sin\!\Big(\!\frac{2k}{Ha}\!\Big)\!\Bigg\}\!+\!\mathcal{O}(\lambda^2)\Bigg\}\;
.\label{Yukawaspectral}\ee
Expanding the trigonometric and integral functions in powers of $\frac{k}{Ha}$ we get
\be &&\hspace{-0.4cm}n_\varphi(t_k,
k)\!=\!-\frac{\lambda}{12\pi^2}\!\Bigg\{\!\!\ln(a)\!-\!\frac{1}{3}\!-\!\frac{a^{-3}}{3}
\!+\!\Big(\!\frac{k}{Ha}\!\Big)^2\!\Bigg[\!\frac{a^2}{5}\!-\!1\!+\!a^{-1}\!\!-\!\frac{a^{-3}}{5}\Bigg]
\!\!-\!\Big(\!\frac{k}{Ha}\!\Big)^4\!\Bigg[\!\frac{3a^4}{70}\!-\!\frac{1}{2}
\!+\!\frac{3a^{-1}}{5}\!-\!\frac{a^{-3}}{7}\Bigg]\nonumber\\
&&\hspace{-0.4cm}+\Big(\!\frac{k}{Ha}\!\Big)^6\!\Bigg[\!\frac{2a^6}{567}
\!+\!\frac{2a^3}{27}\!-\!\frac{2\ln(a)}{9}\!-\!\frac{10}{27}\!+\!\frac{3a^{-1}}{7}\!-\!\frac{11a^{-3}}{81}\Bigg]
\!\!-\!\Big(\!\frac{k}{Ha}\!\Big)^8\!\Bigg[\!\frac{a^8}{5940}
\!+\!\frac{2a^5}{225}\!+\!\frac{2a^3}{45}\!+\!\frac{2a^2}{45}\!-\!\frac{4\ln(a)}{15}\nonumber\\
&&\hspace{5.3cm}-\frac{37}{100}\!+\!\frac{11a^{-1}}{27}\!-\!\frac{67a^{-3}}{495}\Bigg]
\!+\!\mathcal{O}\!\left(\!\Big(\!\frac{k}{Ha}\!\Big)^{10}\right)\!\!\Bigg\}\!+\!\mathcal{O}(\lambda^2)\; .\label{spectrvarphi}
\ee
Analytical result~(\ref{Yukawaspectral}) yields a simple exact expression for the running of the spectral index
\be
&&\hspace{-1.5cm}\alpha_\varphi(t_k, k)\!=\!-\frac{\lambda}{4\pi^2}\frac{1}{\frac{k^2}{H^2}\!+\!a^2}\Bigg\{\!(1\!-\!a)\!\!\left[1\!+\!a\frac{H^2}{k^2}\right]
\!\cos\!\Big(\frac{2k}{H}\!\left(1\!-\!a^{-1}\right)\!\Big)\nonumber\\
&&\hspace{2.5cm}+\frac{H^3}{2k^3}\!\!\left[a^2\!\!-\!(1\!-\!4a\!+\!a^2)\frac{k^2}{H^2}\!+\!\frac{k^4}{H^4}\right]
\!\sin\!\Big(\frac{2k}{H}\!\left(1\!-\!a^{-1}\right)\!\Big)\!\Bigg\}\!+\!\mathcal{O}(\lambda^2)\; ,\label{runAnalytic}
\ee
from which we obtain
\be &&\hspace{-0.5cm}\alpha_\varphi(t_k,
k)\!=\!-\frac{\lambda}{12\pi^2}\!\Bigg\{\!1\!-\!a^{-3}
\!+\!\Big(\!\frac{k}{Ha}\!\Big)^2\!\Bigg[\!\frac{2a^2}{5}\!-\!a^{-1}\!+\!\frac{3a^{-3}}{5}\Bigg]
\!\!-\!\Big(\!\frac{k}{Ha}\!\Big)^4\!\Bigg[\!\frac{6a^4}{35}
\!-\!\frac{3a^{-1}}{5}\!+\!\frac{3a^{-3}}{7}\!\Bigg]\nonumber\\
&&\hspace{-0.5cm}+\Big(\!\frac{k}{Ha}\!\Big)^6\!\Bigg[\!\frac{4a^6}{189}
\!+\!\frac{2a^3}{9}\!-\!\frac{2}{9}\!-\!\frac{3a^{-1}}{7}\!+\!\frac{11a^{-3}}{27}\Bigg]
\!\!-\!\Big(\!\frac{k}{Ha}\!\Big)^8\!\Bigg[\!\frac{2a^8}{1485}
\!+\!\frac{2a^5}{45}\!+\!\frac{2a^3}{15}\!+\!\frac{4a^2}{45}\!-\!\frac{4}{15}\!-\!\frac{11a^{-1}}{27}\nonumber\\
&&\hspace{7.8cm}+\frac{67a^{-3}}{165}\Bigg]\!\!+\!\mathcal{O}\!\left(\!\Big(\!\frac{k}{Ha}\!\Big)^{10}\right)\!\!\Bigg\}
\!+\!\mathcal{O}(\lambda^2)\; .\label{runnvarphi}
\ee
Equations (\ref{powervarphi}), (\ref{spectrvarphi}) and (\ref{runnvarphi}) imply that the power spectrum is red-tilted. Unlike the Yukawa scalar model, where we found that the spectrum is blue-tilted, the amplitudes of fluctuations in the MMC $\lambda\varphi^4$ model grow slightly toward the larger scales. Moreover, in the MMC $\lambda\varphi^4$ model, the one-loop correction in each of the $\Delta^2_\varphi$, $n_\varphi$ and $\alpha_\varphi$ consists of a $k$-independent part and a $k$-dependent part, whereas in the Yukawa scalar model the one-loop corrections have no $k$-independent part. The $k$-independent parts in $\Delta^2_\varphi$ and $n_\varphi$ have logarithmically growing terms. In both models, the $k$-dependent parts approach constants in the late time limit.

\section{Conclusions}
\label{sec:conc}
In the first part of the paper (Secs.~\ref{sec:Model}-\ref{sec:Power}) we considered a MMC scalar $\phi(x;\vec{k})$ which is Yukawa-coupled to a massless Dirac fermion in a locally de Sitter background. We assumed weak Yukawa coupling and applied perturbation theory. In Sec.~\ref{sec:EffectFieldEq} we presented the linearized one-loop Schwinger-Keldish effective field equation for the scalar. We solved it in Sec.~\ref{sec:quantcorrectmodeYukawa} and obtained one-loop mode function~(\ref{g1asSUM}),
\beeq
\mathrm{g}_1(\eta,k)\!=\!\sum_{s=1}^6\mathrm{g}_{1,s}(\eta,k)\; .\nonumber
\eneq
One can get an analytical result for $g_1(\eta,k)$ by adding up $g_{1,s}$ given in Eqs.~(\ref{g1mid}), (\ref{g2mid}), (\ref{g3mid}), (\ref{g4mid}), (\ref{intsing5}) and (\ref{mathcalA})-(\ref{mathcalB}), (\ref{intsing6}) and (\ref{mathcalC})-(\ref{mathcalD}). $\mathrm{g}_1(\eta,k)$ is expressed as a series in powers of $\frac{k}{Ha}$ in Eq.~(\ref{g1pow}). All the terms in the expansion are $k$-dependent and there is no growing term. Hence, $\mathrm{g}_1(\eta,k)$ approaches a constant, for each mode $k$, in the late time limit.

The analytical result for $\mathrm{g}_1(\eta,k)$ can be used to obtain analytical expressions for the one-loop corrected power spectrum $\Delta^2_\phi$, spectral index $n_\phi$ and the running of the spectral index~$\alpha_\phi$. We presented their series expansions in powers of $\frac{k}{Ha}$ in Sec.~\ref{sec:Power}. (See Eqs.~(\ref{Yukpwrspctrm}), (\ref{indexYukawa}) and (\ref{runnYukawa}).) The one loop corrections are all $k$-dependent. Time dependent terms decay in time as inverse powers of the scale factor $a$, hence the one loop corrections approach constants in the late time limit. The $\Delta^2_\phi$ is slightly
blue-tilted; the amplitudes of fluctuations grow slightly
toward the smaller~scales.

In the second part (Sec.~\ref{sec:MMCS}) we applied the computational method used in the first part to MMC $\lambda\varphi^4$ model to get analytical expressions for the one-loop corrected mode function $u_1$ (Eq.~\ref{Phi1}), power spectrum $\Delta^2_\varphi$ (Eq.~\ref{powerwithN}), spectral index $n_\varphi$ (Eq.~\ref{Yukawaspectral}) and running of the spectral index $\alpha_\varphi$ (Eq.~\ref{runAnalytic}) during inflation. Each consists of a $k$-independent part and a $k$-dependent part that approaches a constant---as for the Yukawa scalar---in the late time limit. The $k$-independent parts of the one-loop mode function $u_1$ and the power spectrum $\Delta^2_\varphi$ grow as $\ln^2(a)$ whereas the one-loop spectral index $n_\varphi$ grows as $\ln(a)$. The $k$-independent part of the one-loop running $\alpha^2_\varphi$, on the other hand, freezes to a constant value in the late time limit. The $\Delta^2_\varphi$ is slightly
red-tilted; the amplitudes of fluctuations grow slightly
toward the larger~scales.

\begin{acknowledgements}
I thank Richard P. Woodard for stimulating discussions.
\end{acknowledgements}

\begin{appendix}
\section{Computing the series expansions} \label{App:powerexp}
In this appendix, we give the power and asymptotic series expansions of the cosine and sine integral functions ${\rm ci}(x)$ and ${\rm si}(x)$ that we use throughout the paper.

The power series expansion of function ${\rm ci}(x)$ can be found as follows. By definition,
\be
{\rm ci}(x)\!&=&\!-\!\int_x^\infty\!\! dt\frac{\cos(t)}{t}\!=\!-\!\int_0^\infty\!\!dt\frac{\cos(t)}{t}\!+\!\int_0^x\!\! dt\frac{\cos(t)}{t}\!-\!\int_0^x\!\!\frac{dt}{t}\!+\!\int_0^x\!\!\frac{dt}{t}\nonumber\\
\!&=&\!-\!\int_0^\infty\!\!dt\frac{\cos(t)}{t}\!+\!\int_0^x\!\!\frac{dt}{t}\!+\!\!\int_0^x\!\! dt\frac{\cos(t)\!-\!1}{t}\; .
\ee
Singularities in the first two integrals cancel each other and yield a finite result
\be
{\rm ci}(x)\!=\!\gamma\!+\!\ln(x)\!+\!\!\int_0^x\!\! dt\frac{\cos(t)\!-\!1}{t}\; ,
\ee
where $\gamma\approx0.577$ is the Euler's constant. Power expanding the integrand and integrating it we find
\be
{\rm ci}(x)\!=\!\gamma\!+\!\ln(x)\!+\!\!\sum_{n=1}^\infty
                 \frac{(-1)^{n}x^{2n}}{2n\,(2n)!}\label{cipowerser}\; .
\ee

The power series expansion of function ${\rm si}(x)$ can be found similarly. By definition,
\be
{\rm si}(x)\!=\!-\!\int_x^\infty\!\! dt\frac{\sin(t)}{t}\!=\!-\!\int_0^\infty\!\! dt\frac{\sin(t)}{t}\!+\!\int_0^x\!\! dt\frac{\sin(t)}{t}\!=\!-\frac{\pi}{2}\!+\!\!\int_0^x\!\! dt\frac{\sin(t)}{t}\; .
\ee
Power expanding the integrand and integrating it we find
\be
{\rm si}(x)\!\!&=&\!\!-\frac{\pi}{2}\!+\!\!\sum_{n=1}^\infty
                 \frac{(-1)^{n+1}x^{2n-1}}{(2n\!-\!1)\,(2n\!-\!1)!}\; .\label{sipowerser}
\ee

The asymptotic form of function ${\rm ci}(x)$ for large argument can be found using the definition and integration by parts,
\be
\hspace{-0.5cm}{\rm ci}(x)\!=\!-\!\int_x^\infty\!\! dt\frac{\cos(t)}{t}\!=\!\frac{\sin(x)}{x}\!-\!\int_x^\infty\!\! dt\frac{\sin(t)}{t^2}
\!=\!\frac{\sin(x)}{x}\!-\!\frac{\cos(x)}{x^2}\!+\!2\!\int_x^\infty\!\! dt\frac{\cos(t)}{t^3}\; .
\ee
Continuing the integration by parts yields
\be
\hspace{-0.5cm}{\rm ci}(x)\!=\!\frac{\sin(x)}{x}\!\left(\!1\!-\!\frac{2!}{x^2}
\!+\!\frac{4!}{x^4}\!-\!\frac{6!}{x^6}\dots\right)
\!-\!\frac{\cos(x)}{x}\!\left(\frac{1}{x}\!-\!\frac{3!}{x^3}
\!+\!\frac{5!}{x^5}\!-\!\frac{7!}{x^7}\dots\right)\!\longrightarrow\!\frac{\sin(x)}{x}\; .\label{ciasym}
\ee
The asymptotic form of function ${\rm si}(x)$ for large argument can be found similarly
\be
\hspace{-0.5cm}{\rm si}(x)\!=\!-\!\int_x^\infty\!\! dt\frac{\sin(t)}{t}\!=\!-\frac{\cos(x)}{x}\!+\!\int_x^\infty\!\! dt\frac{\cos(t)}{t^2}
\!=\!-\frac{\cos(x)}{x}\!-\!\frac{\sin(x)}{x^2}\!+\!2\!\int_x^\infty\!\! dt\frac{\sin(t)}{t^3}\; .
\ee
Hence
\be
\hspace{-0.5cm}{\rm si}(x)\!=\!-\frac{\sin(x)}{x}\!\left(\frac{1}{x}\!-\!\frac{3!}{x^3}
\!+\!\frac{5!}{x^5}\!-\!\frac{7!}{x^7}\dots\right)
\!-\!\frac{\cos(x)}{x}\!\left(1\!-\!\frac{2!}{x^2}\!+\!\frac{4!}{x^4}
\!-\!\frac{6!}{x^6}\dots\right)\!\longrightarrow\!-\frac{\cos(x)}{x}\; .\label{siasym}
\ee

\section{Computing the integrals}
In Sec.~\ref{sec:quantcorrectmodeYukawa} we compute the one-loop correction $\mathrm{g}_1(\eta, k)$ to the tree-order (Bunch-Davies) mode function $\mathrm{g}_0(\eta, k)$ by solving the linearized Schwinger-Keldish effective field equation. The $\mathrm{g}_1(\eta, k)$ is obtained by summing up six terms $\mathrm{g}_{1, s}(\eta,k)$ ($s\!=\!1, 2, \ldots, 6$). Computation of each term involves evaluations of several definite integrals. In this appendix, we present the details of the computations. The results of the indefinite integrals are given up to a constant of integration.
\subsection{Computing the $\mathrm{g}_{1,1}(\eta,k)$} \label{App:g11}
Computation of $\mathrm{g}_{1,1}(\eta,k)$ [Eq.~(\ref{g11integrals})] involves integrals
\beeq
\int\!\!da'\!\left[a'^2\!+\!\frac{k^2}{H^2}\right]\!\ln(a')
\!=\!\frac{a'^3}{3}\!\left[\ln(a')\!-\!\frac{1}{3}\right]\!+\!\frac{k^2}{H^2}a'\Big[\!\ln(a')\!-\!1\Big]\; ,\label{intg11int1}
\eneq
and
\be
&&\hspace{-1.3cm}\int\!\!da'\!\left[a'\!-\!\frac{ik}{H}\right]^2\!\!\!e^{\frac{2ik}{Ha'}}\!\ln(a')
\!=\!\frac{a'}{3}e^{\frac{2ik}{Ha'}}\Bigg\{\!a'\!\left[\ln(a')\!-\!\frac{1}{3}\right]\!\!\left[a'\!-\!\frac{2ik}{H}\right]
\!\!+\!\frac{k^2}{H^2}\!\left[\ln(a')\!-\!\frac{7}{3}\right]\!\!\Bigg\}\nonumber\\
&&\hspace{-1.3cm}-\frac{2}{3}\frac{ik^3}{H^3}\Bigg\{\!\frac{\ln^2(a')}{2}\!+\!\!\left[\ln(a')\!-\!\frac{7}{3}\right]
\!\!\Bigg[{\rm ci}\Big(\!\frac{2k}{Ha'}\Big)\!-\!\ln\!\Big(\!\frac{2k}{H}\Big)\!-\!\gamma\!+\!i\!\left[\frac{\pi}{2}\!+\!{\rm si}\Big(\!\frac{2k}{Ha'}\Big)\!\right]\!\Bigg]\!\!-\!\frac{25}{9}\!\Bigg\}\nonumber\\
&&\hspace{6.5cm}+\frac{4}{3a'}\frac{k^4}{H^4}{}_{3}\mathcal{F}_{3}
\!\left(\!\!1, 1, 1; 2, 2, 2; \frac{2ik}{Ha'}\!\right)\!\; ,\label{intg11int2}
\ee
where ${}_{3}\mathcal{F}_{3}
\!\left(1, 1, 1; 2, 2, 2; z\right)$ is a generalized hypergeometric function whose power series expansion
\beeq
{}_{3}\mathcal{F}_{3}
\!\left(1, 1, 1; 2, 2, 2; z\right)\!=\!\sum_{n=0}^\infty\frac{z^n}{(n\!+\!1)^2(n\!+\!1)!}\; .\label{hyp33}
\eneq
Asymptotic form of the hypergeometric function for large argument is
\beeq
{}_{3}\mathcal{F}_{3}
\!\left(1, 1, 1; 2, 2, 2; z\right)\!\rightarrow\!-\frac{1}{z}\left[\frac{\ln^2(z)}{2}
\!+\!\left(\gamma\!+\!i\pi\right)\ln(z)\!+\!\frac{\gamma^2}{2}\!+\!i\gamma\pi\!-\!\frac{5\pi^2}{12}\right]\; .
\eneq

\subsection{Computing the $\mathrm{g}_{1,2}(\eta,k)$} \label{App:g12}
Computation of $\mathrm{g}_{1,2}(\eta,k)$ [Eq.~(\ref{g2ints})] involves two integrals:
\beeq
\int\!\!da'
\Big[1\!+\!\frac{ik}{Ha'}\Big]\!=\!a'\!+\!\frac{ik}{H}\ln(a')\; ,\label{intexpint1}
\eneq
and
\beeq
\int\!\!da'
\Big[1\!-\!\frac{ik}{Ha'}\Big]e^{\frac{2ik}{Ha'}}\!=\!a'e^{\frac{2ik}{Ha'}}
\!-\!\frac{ik}{H}{\rm{Ei}}\Big(\!\frac{2ik}{Ha'}\!\Big)\; ,\label{intexpint2}
\eneq
where the exponential integral function ${\rm{Ei}}(z)$, for pure imaginary arguments, can be expressed~[Eq.~(\ref{Ei})] in terms of cosine and sine integral functions.
\subsection{Computing the $\mathrm{g}_{1,3}(\eta,k)$} \label{App:g13}
The two integrals required to compute the $\mathrm{g}_{1,3}(\eta,k)$ [Eq.~(\ref{g3ints})]  are
\be
\int\!\!da' a'
\Big[1\!+\!\frac{k^2}{H^2a'^2}\Big]\!\!&=&\!\!\frac{a'^2}{2}\!+\!\frac{k^2}{H^2}\ln(a')\; ,\label{g13int1}\\
\int\!\!da' a'
\Big[1\!-\!\frac{ik}{Ha'}\Big]^2e^{\frac{2ik}{Ha'}}\!\!&=&\!\!a'\Big[\frac{a'}{2}\!-\!\frac{ik}{H}\Big]e^{\frac{2ik}{Ha'}}
\!-\!\frac{k^2}{H^2}{\rm{Ei}}\Big(\!\frac{2ik}{Ha'}\!\Big)\; .\label{g13int2}
\ee

\subsection{Computing the $\mathrm{g}_{1,4}(\eta,k)$} \label{App:g14}

Computation of $\mathrm{g}_{1,4}(\eta,k)$ [Eq.~(\ref{g14ints})] involves two types of integrals: (i) the ones that are independent of $n$ and (ii) the ones that depend on $n$. The integrals that are independent of $n$ are easy to evaluate
\be
\int\!\! da'
a' \mathrm{g}_0^*(\eta', k)\!\!&=&\!\!\mathrm{g}_0(0, k)\frac{1}{2}\Bigg\{\!a'\!\left[a'\!+\!\frac{ik}{H}\right]\!e^{-\frac{ik}{Ha'}}
\!-\!\frac{k^2}{H^2}{\rm{Ei}}\Big(\!\!-\!\frac{ik}{Ha'}\!\Big)\!\Bigg\}\; ,\label{intg14int1}\\
\int\!\! da'
a' \mathrm{g}_0(\eta', k)\!\!&=&\!\!\mathrm{g}_0(0, k)\frac{1}{2}\Bigg\{\!a'\!\left[a'\!-\!\frac{ik}{H}\right]\!e^{\frac{ik}{Ha'}}
\!-\!\frac{k^2}{H^2}{\rm{Ei}}\Big(\!\frac{ik}{Ha'}\!\Big)\!\Bigg\}\; .\label{intg14int2}
\ee
The second type of integrals, in Eq.~(\ref{g14ints}), depend on $n$ through the factor
\beeq
\left[1\!-\!\frac{1}{a'}\right]^{n-2}\!=\!\sum_{m=0}^{n-2}\frac{(n\!-\!2)!\,(-1)^m\, a'^{-m}}{(n\!-\!2\!-\!m)!\,m!}\; .
\eneq
The first of the two $n$-dependent integrals is
\be
&&\hspace{-1.8cm}\int\!\!da' \mathrm{g}_0^*(\eta', k)\!\!\left[1\!-\!\frac{1}{a'}\right]^{n-2}
\!\left[a'\!-\!\frac{3}{2}\!+\!\frac{ik}{2Ha'}\right]\!=\!\mathrm{g}_0(0, k)\!\sum_{m=0}^{n-2}\frac{(n\!-\!2)!\,(-1)^m}{(n\!-\!2\!-\!m)!\,m!}\nonumber\\
&&\hspace{5.7cm}\times\!\!\int\!\!\frac{da'}{a^{'m}}\!\left[1\!+\!\frac{ik}{Ha'}\right]
\!e^{-\frac{ik}{Ha'}}\!\left[a'\!-\!\frac{3}{2}\!+\!\frac{ik}{2Ha'}\right]\; ,\label{intg14int3}
\ee
where
\be
&&\hspace{-1.3cm}\int\!\!\frac{da'}{a^{'m}}\!\left[1\!+\!\frac{ik}{Ha'}\right]
\!e^{-\frac{ik}{Ha'}}\!\left[a'\!-\!\frac{3}{2}\!+\!\frac{ik}{2Ha'}\right]\!=\!-\left[\frac{ik}{H}\right]^{-m}
\!\!\Bigg\{\!\frac{ik}{H}\Bigg[\frac{3}{2}\Gamma\Big(m\!-\!1, \frac{ik}{Ha'}\Big)\!+\!\Gamma\Big(m, \frac{ik}{Ha'}\Big)\nonumber\\
&&\hspace{2cm}-\frac{1}{2}\Gamma\Big(m\!+\!1, \frac{ik}{Ha'}\Big)\Bigg]\!\!+\!\frac{k^2}{H^2}\Bigg[\Gamma\Big(m\!-\!2, \frac{ik}{Ha'}\Big)\!+\!\Gamma\Big(m\!-\!1, \frac{ik}{Ha'}\Big)\Bigg]\!\Bigg\}\; .\label{intg14int4}
\ee
The incomplete gamma function
\beeq
\Gamma\left(\alpha, z\right)\!=\!\int_{z}^\infty\!\!dt\, t^{\alpha-1}\, e^{-t} \; .\label{incompgamma}
\eneq
Hence
\beeq
\Gamma\left(0, z\right)\!=\!-{\rm{Ei}}(-z)\; .
\eneq
If $\alpha$ is an integer $n$, we have\beeq
\Gamma\left(n, z\right)\!=\!(n\!-\!1)!\, e^{-z}\!\sum_{m=0}^{n-1}\frac{z^m}{m!}\; .\label{seriesincompgamma}
\eneq
The second $n$-dependent integral in Eq.~(\ref{g14ints})
\be
&&\hspace{-1.5cm}\int\!\!da' \mathrm{g}_0(\eta', k)\!\!\left[1\!-\!\frac{1}{a'}\right]^{n-2}
\!\left[a'\!-\!\frac{3}{2}\!+\!\frac{ik}{2Ha'}\right]\!=\!\mathrm{g}_0(0, k)\!\sum_{m=0}^{n-2}\frac{(n\!-\!2)!\,(-1)^m}{(n\!-\!2\!-\!m)!\,m!}\nonumber\\
&&\hspace{6.2cm}\times\!\!\int\!\!\frac{da'}{a^{'m}}\!\left[1\!-\!\frac{ik}{Ha'}\right]
\!e^{\frac{ik}{Ha'}}\!\left[a'\!-\!\frac{3}{2}\!+\!\frac{ik}{2Ha'}\right]\; ,\label{intg14int5}
\ee
where
\be
&&\hspace{-1cm}\int\!\!\frac{da'}{a^{'m}}\!\left[1\!-\!\frac{ik}{Ha'}\right]
\!e^{\frac{ik}{Ha'}}\!\left[a'\!-\!\frac{3}{2}\!+\!\frac{ik}{2Ha'}\right]\!=\!\left[-\frac{ik}{H}\right]^{-m}
\!\!\Bigg\{\!\frac{ik}{H}\Bigg[\frac{3}{2}\Gamma\Big(m\!-\!1, -\frac{ik}{Ha'}\Big)\!+\!2\Gamma\Big(m, -\frac{ik}{Ha'}\Big)\nonumber\\
&&\hspace{2cm}+\frac{1}{2}\Gamma\Big(m\!+\!1, -\frac{ik}{Ha'}\Big)\Bigg]\!\!-\!\frac{k^2}{H^2}\Bigg[\Gamma\Big(m\!-\!2, -\frac{ik}{Ha'}\Big)\!+\!\Gamma\Big(m\!-\!1, -\frac{ik}{Ha'}\Big)\Bigg]\!\Bigg\}\; .\label{intg14int6}\;
\ee

\subsection{Computing the $\mathrm{g}_{1,5}(\eta,k)$} \label{App:g15}

Computation of $\mathrm{g}_{1,5}(\eta,k)$ [Eq.~(\ref{intsing5})] involves calculations of two terms $\mathcal{A}(\eta, k)$ and $\mathcal{B}(\eta, k)$ defined in Eqs.~(\ref{A}) and (\ref{B}). In computing $\mathcal{A}(\eta, k)$, the integrals that we need are\be
&&\hspace{-1.3cm}\int\!da'\!\left[{a'}^2\!+\!\frac{k^2}{H^2}\right]\!=\!\frac{a'^3}{3}\!+\!\frac{k^2}{H^2}a'
\label{mathcalAint1}\; ,\\
&&\hspace{-1.3cm}\int\!da'\!\left[{a'}^2\!+\!\frac{k^2}{H^2}\right]\!\ln(a')
\!=\!\frac{a'^3}{3}\!\left[\ln(a')\!-\!\frac{1}{3}\right]
\!+\!\frac{k^2}{H^2}a'\Big[\!\ln(a')\!-\!1\Big]\label{mathcalAint2}\; ,\\
&&\hspace{-1.3cm}\int\!da'\!\left[{a'}^2\!+\!\frac{k^2}{H^2}\right]\!{\rm ci}\Big(\!\frac{2k}{Ha'}\!\Big)\!=\!\frac{a'^3}{3}\!\left[{\rm ci}\Big(\!\frac{2k}{Ha'}\!\Big)\!+\!\frac{1}{3}\cos\!\Big(\!\frac{2k}{Ha'}\!\Big)\right]
\!\!-\!\frac{k}{H}\frac{a'^2}{9}\sin\!\Big(\!\frac{2k}{Ha'}\!\Big)\nonumber\\
&&\hspace{3.4cm}+\frac{k^2}{H^2}a'\!\left[{\rm ci}\Big(\!\frac{2k}{Ha'}\!\Big)\!+\!\frac{7}{9}\cos\!\Big(\!\frac{2k}{Ha'}\!\Big)\right]
\!+\!\frac{k^3}{H^3}\frac{14}{9}\!\left[{\rm si}\Big(\!\frac{2k}{Ha'}\!\Big)\!+\!\frac{\pi}{2}\right]\; ,\label{mathcalAint3}
\ee
and
\be
&&\hspace{-1.3cm}\int\!da'\!\left[{a'}^2\!+\!\frac{k^2}{H^2}\right]\!{\rm si}\Big(\!\frac{2k}{Ha'}\!\Big)\!=\!\frac{a'^3}{3}\!\left[{\rm si}\Big(\!\frac{2k}{Ha'}\!\Big)\!+\!\frac{1}{3}\sin\!\Big(\!\frac{2k}{Ha'}\!\Big)\right]
\!\!+\!\frac{k}{H}\frac{a'^2}{9}\cos\!\Big(\!\frac{2k}{Ha'}\!\Big)\nonumber\\
&&\hspace{3.4cm}+\frac{k^2}{H^2}a'\!\left[{\rm si}\Big(\!\frac{2k}{Ha'}\!\Big)\!+\!\frac{7}{9}\sin\!\Big(\!\frac{2k}{Ha'}\!\Big)\right]
\!-\!\frac{k^3}{H^3}\frac{14}{9}{\rm ci}\Big(\!\frac{2k}{Ha'}\!\Big)\; .\label{mathcalAint4}
\ee

In computing $\mathcal{B}(\eta, k)$ [Eq.~(\ref{B})], the first integral that we need is
\beeq
\int\!da'\!\left[a'\!+\!\frac{ik}{H}\right]^2\!e^{-\frac{2ik}{Ha'}}
\!=\!\frac{1}{3}\left\{e^{-\frac{2ik}{Ha'}}a'\!\left[a'^2\!+\!\frac{2ik}{H}a'\!+\!\frac{k^2}{H^2}\right]
\!+\!\frac{2ik^3}{H^3}{\rm{Ei}}\Big(\!\!-\!\frac{2ik}{Ha'}\Big)\Big]\!\right\}\; .\label{mathcalBintfirst}
\eneq
The remaining integral in Eq.~(\ref{B})
\be
\!\int_1^a\!\!\!da'\!\left[a'\!+\!\frac{ik}{H}\right]^2\!e^{-\frac{2ik}{Ha'}}\!\left[\ln(a')\!+\!{\rm ci}\Big(\!\frac{2k}{Ha'}\!\Big)
\!+\!i{\rm si}\Big(\!\frac{2k}{Ha'}\!\Big)\right]\; ,
\ee
can be evaluated by power expanding the terms in the last square brackets
\be
&&\hspace{-0.9cm}\ln(a')\!+\!{\rm ci}\Big(\!\frac{2k}{Ha'}\!\Big)
\!+\!i{\rm si}\Big(\!\frac{2k}{Ha'}\!\Big)\!=\!\ln\!\Big(\!\frac{2k}{H}\Big)\!+\!\gamma\!-\!\frac{i\pi}{2}
\!+\!\!\sum_{n=1}^\infty\!\frac{(-1)^n\!\left(\!\frac{2k}{Ha'}\right)^{2n}}{2n(2n)!}
\!+\!i\!\sum_{n=1}^\infty\!\frac{(-1)^{n+1}\!\left(\!\frac{2k}{Ha'}\right)^{2n-1}}{(2n\!-\!1)(2n\!-\!1)!}\; ,\nonumber
\ee
where we employed Eqs.~(\ref{cipowerser}) and~(\ref{sipowerser}). Hence, the second integral that we need is
\be
&&\hspace{-1.5cm}\int\!da'a'^{-2n}\!\left[a'\!+\!\frac{ik}{H}\right]^2\!e^{-\frac{2ik}{Ha'}}
\!=\!-\frac{2ik^3}{H^3}\!\left[\frac{2ik}{H}\right]^{-2n}\!
\Bigg\{\!4\!\left[\Gamma\Big(2n\!-\!3,\frac{2ik}{Ha'}\Big)
\!+\!\Gamma\Big(2n\!-\!2,\frac{2ik}{Ha'}\Big)\right]\nonumber\\
&&\hspace{9.8cm}+\Gamma\Big(2n\!-\!1,\frac{2ik}{Ha'}\Big)\!\Bigg\}\; .\label{mathcalBinttwo}
\ee
Taking $n\rightarrow n\!-\!\frac{1}{2}$ in Eq.~(\ref{mathcalBinttwo}) yields the result for the final integral that we need
\be
&&\hspace{-1.5cm}\int\!da'a'^{1-2n}\!\left[a'\!+\!\frac{ik}{H}\right]^2\!e^{-\frac{2ik}{Ha'}}\!=\!4\frac{k^4}{H^4}\!\left[\frac{2ik}{H}\right]^{-2n}\!
\Bigg\{\!4\!\left[\Gamma\Big(2n\!-\!4,\frac{2ik}{Ha'}\Big)
\!+\!\Gamma\Big(2n\!-\!3,\frac{2ik}{Ha'}\Big)\right]\nonumber\\
&&\hspace{9.6cm}+\Gamma\Big(2n\!-\!2,\frac{2ik}{Ha'}\Big)\!\Bigg\}\; .\label{mathcalBintthree}
\ee

\subsection{Computing the $\mathrm{g}_{1,6}(\eta,k)$} \label{App:g16}

Computation of $\mathrm{g}_{1,6}(\eta,k)$ [Eq.~(\ref{intsing6})] involves calculations of two terms $\mathcal{C}(\eta, k)$ and $\mathcal{D}(\eta, k)$ defined in Eqs.~(\ref{C}) and (\ref{D}), respectively. In computing $\mathcal{C}(\eta, k)$, the integrals that we need are
\be
&&\hspace{-1.2cm}\int\!\!
da'\!\left[{a'}^2\!+\!\frac{k^2}{H^2}\right]\!\Bigg\{\!{\rm ci}\Big(\frac{2k}{H}\Big[1\!-\!a'^{-1}\Big]\!\Big)
\!-\!\ln\!\Big(\frac{2k}{H}\Big[1\!-\!a'^{-1}\Big]\!\Big)\!-\!\gamma\!\Bigg\}
\!=\!\frac{a'}{3}\Big[a'\!+\!2\Big]\!\sin^2\!\Big(\frac{k}{H}\Big[1\!-\!a'^{-1}\Big]\!\Big)\nonumber\\
&&\hspace{-1.2cm}-\frac{2k}{3H}\Bigg\{\!\frac{a'}{2}\sin\!\Big(\!\frac{2k}{H}\Big[\!1\!-\!a'^{-1}\Big]\!\Big)
\!\!-\!{\rm ci}\Big(\!\frac{2k}{Ha'}\!\Big)\!\!\left[\sin\!\Big(\!\frac{2k}{H}\Big)
\!\!-\!\frac{k}{H}\cos\!\Big(\!\frac{2k}{H}\Big)\!\right]\!\!+\!\!\left[\frac{\pi}{2}\!+\!{\rm si}\Big(\!\frac{2k}{Ha'}\!\Big)\right]\nonumber\\
&&\hspace{-1.2cm}\times\!\!\left[\cos\!\Big(\!\frac{2k}{H}\Big)
\!\!+\!\frac{k}{H}\sin\!\Big(\!\frac{2k}{H}\Big)\!\right]\!\!\Bigg\}
\!+\!\!\left[\frac{{a'}^2}{3}\!+\!\frac{k^2}{H^2}\right]\!a'\!\left[{\rm ci}\Big(\frac{2k}{H}\Big[1\!-\!a'^{-1}\Big]\!\Big)
\!-\!\ln\!\Big(\frac{2k}{H}\Big[1\!-\!a'^{-1}\Big]\!\Big)\!-\!\gamma\right]\nonumber\\
&&\hspace{-1.2cm}+\!\!\left[\!\frac{1}{3}\!+\!\frac{k^2}{H^2}\right]\!\!
\left\{\!{\rm ci}\Big(\!\frac{2k}{Ha'}\!\Big)\!\cos\!\Big(\!\frac{2k}{H}\Big)\!+\!\!\left[\pi\!+\!{\rm si}\Big(\!\frac{2k}{Ha'}\!\Big)\right]\!\sin\!\Big(\!\frac{2k}{H}\Big)\!-\!{\rm ci}\Big(\frac{2k}{H}\Big[\!1\!-\!a'^{-1}\Big]\!\Big)\!+\!\ln(1\!-\!a')\!\right\}\!\!\!\Bigg\}\; ,\label{Cintone}\ee
and
\be
&&\hspace{-0cm}\int\!\!
da'\!\left[{a'}^2\!+\!\frac{k^2}{H^2}\right]\!\Bigg\{\!{\rm si}\Big(\frac{2k}{H}\Big[1\!-\!a'^{-1}\Big]\!\Big)\!+\!\frac{\pi}{2}\!\Bigg\}
\!=\!-\frac{1}{3}\Bigg\{\!a'\!\left[\frac{a'}{2}\!+\!1\right]\! \sin\!\Big(\frac{2k}{H}\Big[1\!-\!a'^{-1}\Big]\!\Big)\nonumber\\
&&\hspace{-0cm}+\!\!\left[\pi\!+\!{\rm si}\Big(\!\frac{2k}{Ha'}\!\Big)\right]\!\cos\!\Big(\!\frac{2k}{H}\Big)\!-\!{\rm ci}\Big(\!\frac{2k}{Ha'}\!\Big)\sin\!\Big(\!\frac{2k}{H}\Big)\!\!-\!\!\left[\frac{\pi}{2}\!+\!{\rm si}\Big(\frac{2k}{H}\Big[1\!-\!a'^{-1}\Big]\!\Big)\right]
\!\!\left[a'^3\!-\!1\!+\!\frac{k^2}{H^2}3(a'\!-\!1)\right]\nonumber\\
&&\hspace{-0cm}+\frac{k}{H}\Bigg\{\!{\rm ci}\Big(\!\frac{2k}{Ha'}\!\Big)\!\!\left[2\cos\!\Big(\!\frac{2k}{H}\Big)
\!-\!\frac{k}{H}\sin\!\Big(\!\frac{2k}{H}\Big)\right]\!\!+\!\!\left[\frac{\pi}{2}\!+\!{\rm si}\Big(\!\frac{2k}{Ha'}\!\Big)\right]\!\!\left[2\sin\!\Big(\!\frac{2k}{H}\Big)
\!+\!\frac{k}{H}\cos\!\Big(\!\frac{2k}{H}\Big)\right]\nonumber\\
&&\hspace{10cm}-a'\cos\!\Big(\frac{2k}{H}\Big[1\!-\!a'^{-1}\Big]\!\Big)\!\Bigg\}\!\Bigg\}\; .\label{Cinttwo}\ee

In computing $\mathcal{D}(\eta, k)$ [Eq.~(\ref{D})] the first integral that we need is
\be
&&\hspace{-1cm}\int\!\!da' \!\left[a'\!-\!\frac{ik}{H}\right]^2\!\!e^{\frac{2ik}{Ha'}}\Bigg\{\!{\rm ci}\Big(\frac{2k}{H}\Big[1\!-\!a'^{-1}\Big]\!\Big)\!-\!\ln\!\Big(\frac{2k}{H}
\Big[1\!-\!a'^{-1}\Big]\!\Big)\!-\!\gamma\!\Bigg\}\nonumber\\
&&\hspace{5cm}\!=\!\sum_{n=1}^\infty \!\frac{(-1)^n\left(\frac{2k}{H}\right)^n}{2n(2n)!}\!\!\!\int\!\!da' \!\left[a'\!-\!\frac{ik}{H}\right]^2\!e^{\frac{2ik}{Ha'}}\Big[1\!-\!a'^{-1}\Big]^{2n}\; .\label{intbinom}
\ee
Using the binomial expansion, the integral in Eq.~(\ref{intbinom}) can be expressed as
\be
&&\hspace{-2cm}\int\!\!da' \!\left[a'\!-\!\frac{ik}{H}\right]^2\!\!e^{\frac{2ik}{Ha'}}\Big[1\!-\!a'^{-1}\Big]^{2n}
\!=\!\sum_{m=0}^{2n} \!\frac{(-1)^m(2n)!}{(2n\!-\!m)!\,m!}\int\!\!da' \!\left[a'\!-\!\frac{ik}{H}\right]^2\!e^{\frac{2ik}{Ha'}}a'^{-m}\; ,\label{g6int1}
\ee
where
\be
&&\hspace{-1cm}\int\!\!da' \!\left[a'\!-\!\frac{ik}{H}\right]^2\!\!e^{\frac{2ik}{Ha'}}a'^{-m}
\!=\!\left[-\frac{2ik}{H}\right]^{3-m}\!\Bigg[\Gamma\Big(m\!-\!3,-\frac{2ik}{Ha'}\Big)
\!+\!\Gamma\Big(m\!-\!2,-\frac{2ik}{Ha'}\Big)\nonumber\\
&&\hspace{9cm}+\frac{1}{4}\Gamma\Big(m\!-\!1,-\frac{2ik}{Ha'}\Big)\Bigg]\; .\label{g6int2}
\ee
Using Eqs.~(\ref{g6int1}) and (\ref{g6int2}) in Eq.~(\ref{intbinom}) we obtain
\be
&&\hspace{-0.5cm}\int\!\!da' \!\left[a'\!-\!\frac{ik}{H}\right]^2\!\!e^{\frac{2ik}{Ha'}}\Bigg\{\!{\rm ci}\Big(\frac{2k}{H}\Big[1\!-\!a'^{-1}\Big]\!\Big)
\!-\!\ln\!\Big(\frac{2k}{H}\Big[1\!-\!a'^{-1}\Big]\!\Big)\!-\!\gamma\!\Bigg\}
\!=\!\!\sum_{n=1}^\infty\!\sum_{m=0}^{2n}\!\frac{(-1)^n i^{1-m}\left(\frac{2k}{H}\right)^{2n-m+3}}{2n(2n\!-\!m)!\,m!}\nonumber\\
&&\hspace{3.5cm}\times\Bigg[\Gamma\Big(m\!-\!3,-\frac{2ik}{Ha'}\Big)
\!+\!\Gamma\Big(m\!-\!2,-\frac{2ik}{Ha'}\Big)\!+\!\frac{1}{4}\Gamma\Big(m\!-\!1,-\frac{2ik}{Ha'}\Big)\Bigg]\; .\label{anahalf1}
\ee
The remaining integral in Eq.~(\ref{D}) is evaluated similarly,
\be
&&\hspace{-1.5cm}\int\!\!da' \!\left[a'\!-\!\frac{ik}{H}\right]^2\!e^{\frac{2ik}{Ha'}}\Bigg\{\!{\rm si}\Big(\frac{2k}{H}\Big[1\!-\!a'^{-1}\Big]\!\Big)\!+\!\frac{\pi}{2}\!\Bigg\}\nonumber\\
&&\hspace{3cm}\!=\!\sum_{n=1}^\infty \!\frac{(-1)^{n+1}\left(\frac{2k}{H}\right)^{2n-1}}{(2n\!-\!1)(2n\!-\!1)!}\!\int\!\!da' \!\left[a'\!-\!\frac{ik}{H}\right]^2\!e^{\frac{2ik}{Ha'}}\Big[1\!-\!a'^{-1}\Big]^{2n-1}\; .\label{intikibinom}
\ee
The integral in Eq.~(\ref{intikibinom}) can be evaluated using the binomial expansion and Eq.~(\ref{g6int2})
\be
&&\hspace{-1cm}\int\!\!da' \!\left[a'\!-\!\frac{ik}{H}\right]^2\!e^{\frac{2ik}{Ha'}}\Big[1\!-\!a'^{-1}\Big]^{2n-1}
\!\!=\!-\!\sum_{m=0}^{2n-1}\!\frac{(2n\!-\!1)!\left(\frac{2ik}{H}\right)^{3-m}}{(2n\!-\!1\!-\!m)!\,m!}\nonumber\\
&&\hspace{2.5cm}\times\!\Bigg[\!\Gamma\Big(m\!-\!3,-\frac{2ik}{Ha'}\Big)
\!+\!\Gamma\Big(m\!-\!2,-\frac{2ik}{Ha'}\Big)\!+\!\frac{1}{4}\Gamma\Big(m\!-\!1,-\frac{2ik}{Ha'}\Big)\Bigg]\; .
\label{g6int3}
\ee
Inserting Eq.~(\ref{g6int3}) into Eq.~(\ref{intikibinom}) yields
\be
&&\hspace{-1.8cm}\int\!\!da' \!\left[a'\!-\!\frac{ik}{H}\right]^2\!e^{\frac{2ik}{Ha'}}\Bigg\{\!{\rm si}\Big(\frac{2k}{H}\Big[1\!-\!a'^{-1}\Big]\!\Big)\!+\!\frac{\pi}{2}\!\Bigg\}
\!=\!\sum_{n=1}^\infty\!\sum_{m=0}^{2n-1}\!\frac{(-1)^{n+1} i^{1-m}\left(\frac{2k}{H}\right)^{2n-m+2}}{(2n\!-\!1)(2n\!-\!1\!-\!m)!\,m!}\nonumber\\
&&\hspace{2cm}\times\!\Bigg[\!\Gamma\Big(m\!-\!3,-\frac{2ik}{Ha'}\Big)
\!+\!\Gamma\Big(m\!-\!2,-\frac{2ik}{Ha'}\Big)\!+\!\frac{1}{4}\Gamma\Big(m\!-\!1,-\frac{2ik}{Ha'}\Big)\Bigg]\; .\label{anahalf2}
\ee

\end{appendix}

\end{document}